\documentclass[prd,twocolumn]{revtex4}
\usepackage{epsfig}
\usepackage{amsmath}
\newcommand{\be}{\begin{equation}}
\newcommand{\ee}{\end{equation}}
\newcommand{\bm}{\begin{mathletters}}
\newcommand{\eem}{\end{mathletters}}
\newcommand{\bes}{\begin{subequations}}
\newcommand{\ees}{\end{subequations}}
\newcommand{\ord}{{\cal O}}
\newcommand{\bi}{\begin{itemize}}
\newcommand{\ei}{\end{itemize}}
\newcommand{\gev}{~{\rm GeV}}

\begin{document}
\title{Brane world unification of quark and lepton masses and
its implication for the masses of the neutrinos}
\author{P.Q. Hung}
\email[]{pqh@virginia.edu}
\affiliation{Dept. of Physics, University of Virginia, \\
382 McCormick Road, P. O. Box 400714, Charlottesville, Virginia 22904-4714, 
USA}
\date{\today}
\begin{abstract}
A TeV-scale scenario is constructed in an attempt to understand the 
relationship between quark and lepton masses. This scenario combines
a model of early (TeV) unification of quarks and leptons with the
physics of large extra dimensions. It demonstrates a relationship
between quark and lepton mass scales at rather ``low'' (TeV) energies
which will be dubbed as {\em early quark-lepton mass unification}.
It also predicts that the masses of the neutrinos are {\em naturally
light} and {\em Dirac}. There is an interesting correlation between
neutrino masses and those of the unconventionally charged fermions
which are present in the early unification model. If these unconventional
fermions were to lie between $200\,GeV$ and $300\,GeV$, the Dirac neutrino
mass scale is predicted to be between 
$\sim 0.07\, eV$ and $\sim 1\,eV$.

\end{abstract}
\maketitle

\section{Introduction}

Are quark and lepton masses related? This question has been addressed
almost thirty years ago in a famous paper by \cite{BEGN} soon after
the concept of Grand Unification (GUT) \cite{GUT} has been put forward. From
this pioneer paper and subsequent works, one learns that quark-lepton
unification at the GUT scale $M_{GUT} \sim 10^{15}-10^{16} \gev$ gives
rise to, for the particular case of $SU(5)$ considered in \cite{BEGN},
the equality of the $\tau$ lepton and bottom quark masses at
$M_{GUT}$. After renomalization-group (RG) evolution down to
low energies, a remarkable ``prediction'' for the $b$ quark mass was 
made, although the complete story was significantly more complicated.
Despite the enormous popularity of GUT, questions started to
arise as to whether or not there are actually structures instead
of simply a ``desert'' between the electroweak scale and $M_{GUT}$.
If so, how would quark and lepton masses be related if they were
to have early unification?

The hope that new physics is lurking somewhere in 
the TeV region has given
rise in the past decade or two to a flurry of activities which resulted
in a rich diversity of topics with a variety of motivations. A common
thread in all of these activities is the prediction of new particles
of one kind or another. It goes without saying that discoveries of
these new particles will vindicate all the efforts put into it. The
present paper will rely on two of such scenarios with a special emphasis
put into the relationship between quark and lepton masses, including
the issues of neutrino mass: Is it Dirac or Majorana? Why is it so small?

Two TeV scenarios which form the focus of this paper are the following:
(1) Early petite unification of quarks and leptons \cite{HBB,BH2,BH3,H1};
(2) The possibility of the existence of extra spatial dimensions,
the mechanism of wave function overlap along an extra compact
spatial dimension \cite{AHS} and its use in the attempt to explain
the smallness of neutrino masses \cite{ref6,H2} and the hierarchy
of quark masses \cite{grossman,HS,HSS}. 

In \cite{HBB}, the Standard Model (SM) with three independent couplings
is merged into a group $G_S\otimes G_{W}$ with two independent
couplings at some scale which is supposed to be in the TeV region. 
The choice of the Pati-Salam $SU(4)_{PS}$ \cite{PS} for $G_S$ was used.
This scenario allowed us to compute $\sin^{2}\theta_{W}(M_{Z}^{2})$ and
to use it to constraint the choices of $G_{W}$. The preferred choice
of \cite{HBB} was the gauge group $SU(4)_{\rm PS}\otimes SU(2)^4$ with
an early unification scale of several hundreds of TeVs. Recent precise
measurements of $\sin^{2}\theta_{W}(M_{Z}^{2})$ coupled with a
surge of interest in TeV scale physics have prompted \cite{BH2} to 
reexamine the petite unification idea. There it was shown that the
petite unification scale is lowered considerably,
to {\em less} than $10\,TeV$, due to the increase
of $\sin^{2}\theta_{W}(M_{Z}^{2})$ as compared with its value of twenty
three years ago. This has the efect of practically ruling out
$SU(4)_{\rm PS}\otimes SU(2)^4$ due to severe problems with the
decay rate for $K_L\to\mu e$ among other things. Two favorite models
emerged: $SU(4)_{\rm PS}\otimes SU(2)^3$ and $SU(4)_{\rm PS}\otimes SU(3)^2$,
both of which nicely and naturally avoid the $K_L\to\mu e$ problem
due pricipally to the existence of new types of fermions. A detailed
analysis of $SU(4)_{\rm PS}\otimes SU(2)^3$ was performed by \cite{BH3},
including a two-loop renormalization group (RG) analysis and a
discussion of the physics of the new unconventional fermions. Early 
unification in this model takes place at a mass scale 
$M=\ord (1-2~{\rm TeV})$.

On another front, Ref. \cite{H2} has constructed a model which made
use of the mechanism of wave function overlap along an extra compact
dimension to ``explain'' the smallness of {\em Dirac} neutrino masses.
An $SU(2)_R$ symmetry was assumed and was subsequently spontaneously
broken, giving rise to a phenomenon in which one member of the
right-handed doublet has a {\em narrow} wave function, while the
other member acquires a {\em broad} wave function (both localized at the same
point along the extra dimension). The overlap of the
wave function of the left-handed doublet with the wave functions
of the right-handed fields gives rise to the splitting between the effective
four-dimensional Yukawa couplings (and eventually 
between the masses) of neutral
and charged leptons or of the up and down quarks. This splitting can be large
or small depending on the separation $d$ (along the extra dimension)
between the wave functions for left-handed and right-handed fields
as demonstrated in \cite{H2}. It was further noticed that there is
a deep connection between the separation $d_l$ for the lepton sector
and the separation $d_q$ for the quark sector, giving rise to a
relationship between quark and lepton masses, a common feature
in Grand Unified Theories (GUT). 

At this point, one might ask about the distinction between the
present scenario and a possible attempt to incorporate a GUT
scenario for the masses in the context of Large Extra Dimensions.
First, it is fair to say that, in order to achieve Grand Unification
above the compactification scale, some rather strong assumption has
to be made about the behaviour of the running couplings, namely
a power-law running. Because of this dynamical assumption,
the running masses used in extrapolating the values at the GUT
scale to low energies will also suffer from large uncertainties. This
is very unlike the logarithmic behaviour used in \cite{BEGN}.
In our case, quark-lepton unification is achieved at a scale
comparable to the compactification scale and the predictions
made there can be extrapolated down to the Z-mass using
familiar renormalization group techniques. In fact, since
the quark-lepton unification scale is an order of magnitude
or so larger than the Z-mass, there will not be much ``running''.

The plan of the paper will be as follows. First, we present 
a brief review
of the essential elements that go into the wave function
overlap scenario in extra dimensions. We then briefly review
the ideas of early quark-lepton unification with a special
emphasis on the group structure and fermion representations. We
then show how one can connect these two ideas to relate the
overall mass scales in the mass matrices of the quark sector
to those of the lepton sectors. We finish with a numerical
illustration of those results along with their physical
implications, including neutrino masses. We will present
predictions for {\em Dirac} neutrino masses. Whether or not
Majorana neutrino masses are needed is a question which
depends on the predicted values for {\em Dirac} neutrino masses.
We will show a correlation between the masses of the neutrinos
and those of the unconventional fermions which are present
in the early unification model. If the latter fermions are
required to have a mass between the electroweak scale and
approximately $1\,TeV$, it is shown that the Dirac neutrino
masses are {\em too small} for the see-saw mechanism 
\cite{seesaw} to
provide the bulk of neutrino masses if, as it is natural
to assume, the Majorana scale is of the order of the
early unifications scale. It is also shown that if the
mass of the unconventional fermions is taken to lie between
$200\,GeV$ and $300\,GeV$, the range of the Dirac neutrino mass
is found to be between $\sim 0.07\, eV$ and $\sim 1\,eV$.
In fact, there is a recent interest
in the possibility that the neutrino mass might be either
mostly or pure Dirac and there are questions
about the popular see-saw mechanism itself \cite{smirnov}.

\section{Extra dimension, Early quark-lepton unification and
mass relationship}

Two TeV-scale scenarios are briefly summarized below with the
purpose of exposing their common threads and ultimately
combining them in order to obtain an understanding of
the possible relationship between quark and lepton masses
and the smallness of the neutrino masses.

\subsection{Effective Yukawa couplings in models with extra dimensions}
\label{LED}

In its simplest version, an effective Yukawa coupling (which would be
proportional to the mass of the fermion) is defined, in four dimensions,
as proportional to the size of the wave function overlap between
left-handed and right-handed fermions along a compact fifth (spatial)
dimension \cite{AHS}. Among the many applications of this idea, one
can cite for example the attempts to give an explanation for
the smallness of the neutrino masses \cite{ref6,H2}. One can either
{\em arbitrarily} choose the locations, along the extra dimension, of the
localized wave functions for the left-handed and right-handed neutrinos
in such a way that the overlap is tiny, or one can try to build a model
in which the tiny overlap comes out more or less naturally as 
Ref. \cite{H2} had done. In \cite{H2}, the size of the neutrino overlap
came out small {\em as compared with} the size of the charged lepton
overlap. A brief review of how this happens as described in Ref. \cite{H2}
will be given below. The main point of these works is that the
four-dimensional effective Yukawa couplings can be {\em small} even if
the fundamental (four-dimensional) Yukawa coupling is of order {\em unity}.

Let us start with one extra spatial dimension $y$ compactified 
on an orbifold $S_{1}/Z_{2}$ and having a length $L$. Let us, as
an example, take a lepton $SU(2)_L$ doublet,  
$L^{\{L\}}(x,y)$, and another lepton
$SU(2)_R$ doublet, $L^{\{R\}}(x,y)$, where the superscripts
refer to the groups respectively.
Since a fermion in five dimensions is a Dirac
fermion, it will have both chiralities (left and right-handed)
under four dimensions i.e. $\psi = 
(\psi_L+ \psi_R)$, where $\psi_{L,R} = P_{L,R}\psi$, with
$P_{L,R} = (1\mp \gamma_5)/2$ being the usual
four-dimensional chiral projection operator.
The notations that were used in \cite{H2} and here will be as follows.
For the $SU(2)_L$ doublet, we use $L^{\{L\}}(x,y) = (l^{\{L\}}_{L}+
l^{\{L\}}_{R})$, while for the $SU(2)_R$ doublet, we use
$L^{\{R\}}(x,y) = (l^{\{R\}}_{L}+l^{\{R\}}_{R})$. One can choose the
$Z_2$ parity for these fields such that the only zero modes are
$l^{0,\{L\}}_{L}(x,y) = l_{L}(x)\xi_{L}(y)$ and
$l^{0,\{R\}}_{R}(x,y) = l_{R}(x)\xi_{R}(y)$. With the introduction
of the appropriate background scalar fields, these zero modes
can be localized at some points along $y$. The effective
Yukawa coupling (which would determine the mass of the fermion)
is then proportional to the overlap between $\xi_{L}(y)$ and
$\xi_{R}(y)$. (Let us recall that $\xi_{L}(y)$ and
$\xi_{R}(y)$ are doublets.) The main focus of \cite{H2} was the
construction of a model for the $SU(2)_R$ doublet $\xi_{R}(y)$.
This construction will be repeated below but a few words might be
illuminating here. The wave functions for the up and down members
of $\xi_{R}(y)$, although localized at the same point along $y$,
have very different shapes: one which is wide and the other which is
narrow. It is this disparity in shapes of the ``right-handed'' wave 
functions that, when overlapping with the common ``left-handed''
wave function, gives rise to the hierarchy in mass among up and
down members of the doublet.

For the sake of clarity, a review of the model of \cite{H2} is
warranted here. Since the main object is the construction of
$\xi_{R}(y)$, one will concentrate on $L^{\{R\}}$. A summary
of the main results of \cite{H2} can now be given. First, the
localization of $\xi_{R}(y)$ can be achieved by a coupling
of $L^{\{R\}}$ to a background scalar field which develops
a kink solution along $y$. Two background scalar fields are needed in
this scenario: a {\em singlet} field $\phi_{S}$ whose kink solution
localizes the wave functions of both members of $\xi_{R}(y)$ at the 
same location while keeping their shapes identical, and a {\em triplet}
$\Phi_{T}= \vec{\phi}_{T}.\frac{\vec{\tau}}{2}$ whose kink solution
is responsible for drastically changing the shapes of the
two wave functions while keeping the localization points the same.
(As mentioned in \cite{H2}, these background scalars are 
chosen to be odd under $Z_2$ so that they do not have zero modes.)
Below is how it works.

The minimum energy solutions used in \cite{H2} are as follows
\begin{equation}
\label{PhiT}
\langle \Phi_{T} \rangle = \langle\phi_{T}^{3}\rangle
\tau_{3}/2 =
\left(\begin{array}{cc}
h_{T}(y)&0 \\
0&-h_{T}(y)
\end{array}
\right)\ \, ,
\end{equation}
and
\begin{equation}
\label{phis}
\langle \phi_{S} \rangle = h_{S}(y) \, ,
\end{equation}
where generically $h(y)=v\,\tanh (\mu y)$, with $\mu =\sqrt{\lambda/2}
v$ being typically the ``thickness'' of the domain wall.
Coupled with the Yukawa coupling
${\cal L}_{Y2}= f_{T}^{(l)}\bar{L}^{\{R\}}\Phi_{T}\,L^{\{R\}} +
f_{S}^{(l)}\bar{L}^{\{R\}}\phi_{S}\,L^{\{R\}}$, one obtains the following
equations for $\xi_{R}(y)$:
\begin{mathletters}
\begin{equation}
\label{nu}
\partial_{y}\xi^{\nu}_{R}(y)+(f_{S}^{(l)}h_{S}(y) +
f_{T}^{(l)}h_{T}(y))\xi^{\nu}_{R}(y) = 0\,,
\end{equation}
\begin{equation}
\label{e}
\partial_{y}\xi^{e}_{R}(y)+(f_{S}^{(l)}h_{S}(y)
-f_{T}^{(l)}h_{T}(y))\xi^{e}_{R}(y) = 0\,,
\end{equation}
\end{mathletters}
The solutions
for the up and down members of $\xi_{R}(y)$ (which will have the
superscripts $\nu$ and $e$ respectively) are given as
\begin{equation}
\label{xinu}
\xi_{R}^{\nu,e}(y)= k_{\nu,e}\exp(-\int_{0}^{y}dy^{\prime} 
(f_{S}^{(l)} h_{S}(y^{\prime})\pm f_{T}^{(l)} h_{T}(y^{\prime}))\,,
\end{equation}
where $k_{\nu,e}$ are normalization factors. The immediate implication
of Eq. \ref{xinu} can be seen as follows. Using $h(y)=v\,\tanh (\mu y)$
in Eq. \ref{xinu}, one obtains
\begin{equation}
\label{xinu2}
\xi_{R}^{\nu,e}(y) = k_{\nu,e} e^{-(C_{S} \ln(\cosh(\mu_{S}y)) \pm
C_{T} \ln(\cosh(\mu_{T}y)))} \, ,
\end{equation}
where $C_{S,T} = f_{S,T}/(\lambda_{S,T}/2)^{1/2}$. If the parameters
of the two scalar potentials are such that 
$C_{S} \ln(\cosh(\mu_{S}y)) \approx C_{T} \ln(\cosh(\mu_{T}y))$,
one can immediately see that $\xi_{R}^{\nu}(y)$ is {\em narrow}
while $\xi_{R}^{e}(y)$ is {\em broad}. When these wave functions
overlap with the left-handed wave function (common for both
$\nu$ and $e$), one can observe a {\em large} disparity between
the two effective Yukawa couplings. A crucial quantity which
enters this hierarchy in Yukawa couplings is the {\em separation}
between the left-handed wave function and the two right-handed
wave functions (localized at the same point along $y$) and
which was denoted by $\Delta y^{(l)}$ in \cite{H2}.

The model described above has been
espoused in Ref. \cite{H2} as a mechanism for naturally small
Dirac neutrino masses. Furthermore, using the same wave function
profiles for the right-handed quarks, an interesting connection
between quark and lepton mass hierarchies was noticed in \cite{H2}.
Basically, it was a connection between $\Delta y^{(l)}$
and $\Delta y^{(q)}$.
Possible symmetry reasons for this connection were left open
in \cite{H2}. It is the purpose of this paper to elucidate
the relationship between quark and lepton mass hierarchies
by considering explicitely a model of TeV-scale quark-lepton
unification \cite{HBB,BH2,BH3}. To set the stage for that
discussion, a brief summary of the early unification model
is presented below.

\subsection{Early TeV-scale quark-lepton unification}
\label{PUT}

The model that was presented in \cite{BH2} and discussed in detail
in \cite{BH3} is based on the gauge group

\be
\label{put1}
G_{PUT}=SU(4)_{\rm PS}\otimes SU(2)_L \otimes SU(2)_R \otimes SU(2)_H \,.
\ee
This group is characterized by {\em two} independent gauge couplings:
$g_S$ for $SU(4)_{\rm PS}$ and $g_W$ for 
$SU(2)_L \otimes SU(2)_R \otimes SU(2)_H$, where a permutation symmetry
is assumed among the three $SU(2)$'s. $G_{PUT}$ is assumed to be
broken down to the Standard Model in two steps, namely
\begin{equation}
\label{pattern}
G_{\rm PUT} \stackrel{\textstyle M}{\longrightarrow} G_1 
\stackrel{\textstyle \tilde{M}}{\longrightarrow} G_2
\stackrel{\textstyle M_Z}{\longrightarrow} SU(3)_c \otimes U(1)_{EM} ,
\end{equation}
where
\begin{equation}
\label{G1}
G_1 = SU(3)_{c} \otimes U(1)_S \otimes
SU(2)_{L}\otimes SU(2)_{R}\otimes SU(2)_{H}                        \, ,
\end {equation}
and
\begin{equation}
\label{G2}
G_2 = SU(3)_{c} \otimes SU(2)_{L} \otimes U(1)_{Y}\,.
\end {equation}

In this scheme, quarks and leptons, which are generic terms
for color triplets and color singlets respectively, are grouped
into quartets of $SU(4)_{PS}$. The scale of such a quark-lepton
unification is denoted by $M$ as seen above. In contrast with
GUT where such a unification occurs close to the Planck scale,
it has been shown in \cite{BH2}, and particularly in \cite{BH3},
that $M \leq 2\,TeV$. That such a low scale of unification can
be achieved is a distinctive feature of this model. A sketch of
the arguments is presented below.

At this point, it is worth noticing that, if the scale(s) of
extra dimensions is comparable with the Petite Unification scale,
physics which are related to the breaking of Petite Unification
can be extrapolated to ``low energies'' with little, if any,
uncertainties coming from physics beyond the compactification
scale. This is in contrast with a typical GUT scenario embedded in
large extra dimensions since its
scale which would normally lie
above the compactification scale. As a consequence, there are
large uncertainties associated with the extrapolation of
``GUT'' physics down to the Z-mass for example.

The main idea of Petite Unification has to do with the assumption
that the SM, with three independent couplings: $g_3$, $g_2$ and $g_1$,
is merged into the PUT group $G_S \otimes G_W$ which is
characterized by two independent couplings: $g_S$ and $g_W$. As
a result, one can compute $\sin^{2}\theta_{W}(M_{Z}^{2})$ as
a function of the PUT unification scale $M$ as shown in \cite{HBB,BH2}.
The highly precise value of $\sin^{2}\theta_{W}(M_{Z}^{2})=0.23113(15)$,
along with the requirement that $M \leq 10\,TeV$,
allows us to severely restrict the choices of $G_W$, resulting in
the preferred model mentioned at the beginning of this section. (Two
other models were also found: $SU(4)_{PS} \otimes SU(2)^4$ and
$SU(4)_{PS} \otimes SU(3)^2$, with the former being, in some sense,
ruled out due to severe problems with the process $K_L \rightarrow
\mu e$ unless some exotic mechanisms are invoked, 
for example an embedding of the
model into five dimensions with the gauge symmetry breaking 
accomplished by orbifold boundary conditions \cite{CHP}.)

Two crucial elements in the computation of
$\sin^{2}\theta_{W}(M_{Z}^{2})$ are the group theoretical factor
$\sin^{2}\theta_{W}^{0}$ ($=1/3$ for $G_W = SU(2)^3$) and the
factor $C_S$ which appears in the expression
$Q=T_{3L}+T_Y=Q_W+C_S T_{15}$, where 
$Q_W$ is the "weak" charge corresponding to the group $G_W$, and
$T_{15}$ is the unbroken diagonal generator of the 
$SU(4)_{\rm PS}$. The value of $C_S$ depends on how
quarks and leptons transform under 
$SU(4)_{\rm PS}\otimes SU(2)_L \otimes SU(2)_R \otimes SU(2)_H$. 
For instance, $C_S = \sqrt{2/3}$ if fermions transform
as $(4,2,1,1)$ for example, while $C_S = \sqrt{8/3}$ if they
transform as $(4,2,1,2)$ or $(4,1,2,2)$.
This is shown in details in \cite{HBB,BH2}. Since
$\sin^{2}\theta_{W}(M_{Z}^{2})= (1/3)(1-0.067\,C_{S}^2-\log terms)$
(see \cite{BH2})
and coupled with the requirement that $M \leq 10\,TeV$ (which makes
for little ``running'' between $M$ and the electroweak scale, and
hence small $\log$ terms), it was found \cite{BH2} that the
only acceptable fermion representations are the ones for which
$C_S = \sqrt{8/3}$. Using this value for $C_S$ \cite{BH2,BH3},
a detailed computation of 
$\sin^{2}\theta_{W}(M_{Z}^{2})$, up to two loops \cite{BH3}, 
determines the Petite Unification scale to be less than $2\,TeV$.
This fermion content will be the one that will be used in this paper.
For the sake of clarity, an explicit description of the fermions
of the model is presented below.

Under $SU(4)_{\rm PS}\otimes SU(2)_L \otimes SU(2)_R \otimes SU(2)_H$,
the fermions transform as
\begin{equation}\label{REP1}
\Psi_L=(4,2,1,2)_L=
\left(\begin{array}{cc}
(d^c(1/3),\tilde U(4/3)) & (\tilde l_u(-1),\nu(0)) \\
(u^c(-2/3),\tilde D(1/3)) & (\tilde l_d(-2),l(-1))
\end{array}\right)_L
\end{equation}

\begin{equation}\label{REP2}
\Psi_R=(4,1,2,2)_R=
\left(\begin{array}{cc}
(d^c(1/3),\tilde U(4/3)) & (\tilde l_u(-1),\nu(0)) \\
(u^c(-2/3),\tilde D(1/3)) & (\tilde l_d(-2),l(-1))
\end{array}\right)_R
\end{equation}
As one can see, this model contains, besides conventionally charged 
fermions, {\em unconventional} fermions with charges up to
$4/3$ for the quarks and down to $-2$ for the leptons.

To understand the notations in Eqs. (\ref{REP1},\ref{REP2}), one
notices the following conventions.

\bi

\item $SU(2)_{L,R}$ doublets:

\be\label{LR}
\left(
\begin{array}{c}
d^{c}(1/3)\\ 
u^{c}(-2/3)
\end{array}
\right)_{L,R},
\qquad
\left(
\begin{array}{c}
\nu(0) \\ 
l(-1)
\end{array}
\right)_{L,R},\nonumber
\ee
\be
\left(
\begin{array}{c}
\tilde{U}(4/3)\\ 
\tilde{D}(1/3)
\end{array}
\right)_{L,R}, 
\qquad
\left(
\begin{array}{c}
\tilde{l}_{u}(-1)\\ 
\tilde{l}_{d}(-2)
\end{array}
\right)_{L,R}
\ee

\item $SU(2)_{H}$ doublets:

\be\label{H}
\left(
\begin{array}{c}
\tilde{U}(4/3)\\ 
d^{c}(1/3)
\end{array}
\right)_{L,R},
\qquad
\left(
\begin{array}{c}
\nu(0) \\ 
\tilde{l}_{u}(-1)
\end{array}
\right)_{L,R},\nonumber
\ee
\be
\left(
\begin{array}{c}
\tilde{D}(1/3)\\ 
u^{c}(-2/3)
\end{array}
\right)_{L,R}, 
\qquad
\left(
\begin{array}{c}
l(-1)\\ 
\tilde{l}_{d}(-2)
\end{array}
\right)_{L,R}
\ee

\item $SU(4)_{PS}$ quartets:

\be\label{LR}
\left(
\begin{array}{c}
d^{c}(1/3)\\ 
\tilde{l}_{u}(-1)
\end{array}
\right)_{L,R},
\qquad
\left(
\begin{array}{c}
\tilde{D}(1/3)\\ 
l(-1)
\end{array}
\right)_{L,R},\nonumber
\ee
\be
\left(
\begin{array}{c}
\tilde{U}(4/3)\\ 
\nu(0) 
\end{array}
\right)_{L,R}, 
\qquad
\left(
\begin{array}{c}
u^{c}(-2/3)\\ 
\tilde{l}_{d}(-2)
\end{array}
\right)_{L,R}
\ee

\ei
Note that due the particular nature of the fermion representation
in this model, it is
\be
\label{doublet}
\left(
\begin{array}{c}
d^{c}(1/3)\\ 
u^{c}(-2/3)
\end{array}
\right)_{L,R} =
i \tau_2
\left(
\begin{array}{c}
u(2/3)\\ 
d(-1/3)
\end{array}
\right)^{*}_{L,R}
\ee
which appears instead of the more familiar-looking $(u(2/3),d(-1/3))$.

As emphasized in \cite{BH2,BH3}, a nice feature of this model is the
absence of tree-level flavor-changing neutral currents (FCNC)
because the $SU(2)_H$ and $SU(4)_{PS}$ transitions only connect
conventional to unconventional fermions as can be seen above. A
process such as $K_L \rightarrow \mu e$ can only occur at one loop
and can easily be made to obey the experimental upper bound.

Since the natural scales of the scenarios described in Sections
\ref{LED} and \ref{PUT} are both in the TeV range, it is worthwhile
to see if a ``marriage'' of some sorts can be made between
these two scenarios and if, as a result, some light can be shed
concerning the relationship between quark and lepton masses and
the smallness of the neutrino masses.

\subsection{Connection between the scales of quark and lepton
masses}
\label{connect}

If ``quarks'' and ``leptons'' (in the generic sense as discussed
above) can be unified at the TeV scale, there
is a good possibility that whatever gives rise to their masses
will also determine the relationship between their mass scales.
We will show below that such a possibility does exist within
the framework of the two scenarios described above.

The basic model used in this paper is 
$SU(4)_{\rm PS}\otimes SU(2)_L \otimes SU(2)_R \otimes SU(2)_H$.
As stated above, this group spontaneously breaks down to
$SU(2)_L \otimes U(1)_Y$ and then to $U(1)_{em}$ at the scales
$M \sim few TeV's$ and $v \sim 250 GeV$ respectively.
Upon embedding this model in five dimensions, with the fifth
dimension $y$ compactified on an $S_1/Z_2$ orbifold, it is shown
below that the following features occur: 1) The breaking
of $SU(4)_{PS}$ splits the positions, along $y$, of wave functions
of the zero modes of ``quarks'' and ``leptons'';
2) The breaking
of $SU(2)_R$ gives rise to two vastly diferent profiles for
the wave functions of the ``right-handed''
zero modes; 3) Since a $SU(2)_H$ doublet groups together
a conventional quark (or lepton) with an unconventional one,
the breaking of $SU(2)_H$ splits the locations, along $y$,
of the wave functions of the conventional fermions relative
to those of the unconventional ones; 4) And finally, the
breaking of $SU(2)_L \otimes U(1)_Y$ provides a mass scale
for all the fermions. As we have explained in the previous
section, a crucial quantity  that appears in the hierarchy
of masses among the up and down members of an $SU(2)_L$ 
doublet is the separation along $y$ between the wave function
of the left-handed doublet and that of the right-handed fields,
namely $\Delta y^{(l,q)}$. We will show below that points
\# 1, 2 and 3 help establish a relationship between
$\Delta y^{(l)}$ and $\Delta y^{(q)}$.

In the construction of the model, one important point we
would like to stress is the following. The model
$SU(4)_{\rm PS}\otimes SU(2)_L \otimes SU(2)_R \otimes SU(2)_H$
contains unconventional quarks and leptons which were
assumed to be heavy enough to escape detection. The fate
of these fermions were well described in Ref. \cite{BH2}. For
the purpose of this paper, we will simply require that 
{\em all unconventional fermions} are {\em heavy}. This will be
one constraint which will be used below.

\subsubsection{$SU(4)_{\rm PS}\otimes SU(2)_L \otimes SU(2)_R \otimes SU(2)_H$
in five dimensions}
\label{5d}

The first step one would like to do is to embed
$SU(4)_{\rm PS}\otimes SU(2)_L \otimes SU(2)_R \otimes SU(2)_H$
in five dimensions.
Let us first denote, in five dimensions, the fermions presented in 
Section \ref{PUT} by
\be
\label{psil}
\Psi^{\{L\}}(x,y) = (4,2,1,2) \,,
\ee
\be
\label{psir}
\Psi^{\{R\}}(x,y) = (4,1,2,2) \,.
\ee
Let us recall that, in five dimensions, these are four-component 
Dirac fields, i.e. they have both left- and right-handed
components. The superscripts $\{L\}$ and $\{R\}$ are used for
two reasons: (a) to denote the transformation under $SU(2)_L$
or $SU(2)_R$; and (b) to show that the surviving zero modes
are related to these fields.
By choosing the appropriate $Z_2$ parity for these
fields, the zero modes of $\Psi^{\{L\}}(x,y)$ and
$\Psi^{\{R\}}(x,y)$ are 
\bes
\label{zero}
\be
\Psi_{L}(x,y) = \psi_{L}(x)\, \xi_{L}(y) \,,
\ee
\label{zero}
\be
\Psi_{R}(x,y) = \psi_{R}(x)\, \xi_{R}(y) \,,
\ee
\ees
respectively.

We wish to localize $\xi_{L}(y)$ and
$\xi_{R}(y)$ along $y$. This is accomplished by coupling
these fields to some background scalar fields. To see
the group representations of these scalar fields, we 
consider the following bilinears:
\be
\label{bil}
\bar{\Psi}^{\{L\}}(x,y)\Psi^{\{L\}}(x,y)=(1+15,1+3,1,1+3) \,,
\ee
\be
\label{bir}
\bar{\Psi}^{\{R\}}(x,y)\Psi^{\{R\}}(x,y)=(1+15,1,1+3,1+3) \,.
\ee

From Eq. (\ref{bil}), one can see that some possible scalar fields which
can couple to these fermions would transform like
$(1,1,1,1)$, $(15,1,1,1)$, $(15,1,1,3)$, $(15,1,3,1)$, etc..
We will show step-by-step below how these scalar fields help
establish the link between $\Delta y^{(l)}$ and $\Delta y^{(q)}$.
We will successively invoke one scalar at a time and show
how it modifies the behaviour of the fermion zero modes.

As we shall see, the scalar fields which are needed for our scenario are 
the following:
\bes
\label{scalars}
\be
\Phi_{S_1,S_2} = (1,1,1,1) \,
\ee
\be
\Sigma = (15,1,1,1) \,,
\ee
\be
\Phi_R = (1,1,3,1) \,,
\ee
\be
\Phi_H = (15,1,1,3) \,.
\ee
\ees

\subsubsection{The role of
the singlet scalar fields $\Phi_{S_1,S_2} = (1,1,1,1)$}
\label{singlet}

The Yukawa coupling of this field with the fermions takes the form
\be
\label{yuk1}
{\cal L}_{Y1} = f_{S}(\bar{\Psi}^{\{L\}}\Psi^{\{L\}}
+\bar{\Psi}^{\{R\}}\Psi^{\{R\}})\Phi_{S_1} \,,
\ee
with $f_S >0$. We assume a kink solution for $\Phi_{S_1}$
\be
\label{kink1}
\langle \Phi_{S_1} \rangle = h_{S}(y) \,,
\ee
which localizes {\em all} fermions at the same point $y=0$ 
along $y$. In fact,the
equation governing the zero modes, at this stage, is

In order to be more general, we will allow the fermions to be localized,
still at this stage, at some common arbitray point which might
be different from the origin. The most economical scenario is
one in which the ``left-handed'' fermions are localized at that other point.
This can be accomplished by the following coupling:
\be
\label{yuk1p}
{\cal L}_{Y1p} = -f_{S'}\Phi_{S_2}\, \bar{\Psi}^{\{L\}}\Psi^{\{L\}} \,,
\ee
where $f_{S'}>0$ and the negative sign in front of it is
an arbitrary choice.
Assuming 
\be
\label{phis2}
\langle \Phi_{S_2} \rangle = \delta \,,
\ee
we obtain the following equations for the zero modes:
\bes
\label{zeroeq1}
\be
\partial_{y}\xi_{L}(y)+\{f_{S}\,h_{S}(y)-
f_{S'}\,\delta\}\xi_{L}(y) = 0\,.
\ee
\be
\partial_{y}\xi_{R}(y)+\{f_{S}\,h_{S}(y)\}\xi_{R}(y) = 0\,.
\ee
\ees

We will present below two scenarios.
\bi

\item Scenario I:

\be
\delta = 0 \,.
\ee

\item Scenario II:

\be
\delta \neq 0 \,.
\ee

\ei

{\em At this stage}, Scenario I implies that all (left and right-handed)
fermions are localized at the origin. Scenario II implies that
the right-handed ones are localized at the origin while the
left-handed ones are localized at a common point away from the
origin. As we shall see in the last section, it will be
Scenario II with $\delta \neq 0$ that is favored phenomenologically.

It is clear that this is not the end of the story because the
effective Yukawa couplings to an $SU(2)_L$-doublet Higgs field,
which depend on the overlap of the left and right wave
functions, would be universal for all fermions, a clearly
undesirable feature. We therefore have to split the various
wave functions along $y$. To do this, one has to invoke
scalars which transform non-trivially under
$SU(4)_{\rm PS}\otimes SU(2)_L \otimes SU(2)_R \otimes SU(2)_H$.
This is what we will proceed to do next.

\subsubsection{The roles of $\Sigma = (15,1,1,1)$ and $\Phi_H = (15,1,1,3)$}
\label{phih}

From the group representations of $\Sigma = (15,1,1,1)$ and 
$\Phi_H = (15,1,1,3)$, one can
see that, in principle, both $\Psi^{\{L\}}$ and $\Psi^{\{R\}}$
can couple to $\Sigma$ and $\Phi_H$. However, for reasons of economy, we shall
see that it is sufficient to couple $\Psi^{\{L\}}$ to
$\Phi_H$.

We now concentrate on $\xi_L(y)$. As it is mentioned above, one
has to differentiate the unconventional fermions from the conventional
ones as well as the ``quarks'' from the ``leptons''. 
Let us remind ourselves that the conventional and unconventional
fermions are grouped into $SU(2)_H$ doublets as shown in Eq. (\ref{H}).
To differentiate the aforementioned fermions, we
need to break both $SU(4)_{PS}$ and $SU(2)_H$. This is
accomplished by the use of $\Phi_H = (15,1,1,3)$ and of
$\Sigma = (15,1,1,1)$. The Yukawa
interaction between $\Psi^{\{L\}}$, $\Phi_H$ and $\Sigma = (15,1,1,1)$ is given by
\be
\label{yuk2}
{\cal L}_{Y2} = 
\bar{\Psi}^{\{L\}}\,(f_{H}\,\Phi_{H}+ f_{\Sigma} \Sigma)\,\Psi^{\{L\}} \,,
\ee
with $f_H, f_{\Sigma} >0$.

We will assume a vacuum expectation value for $\Phi_H$  and $\Sigma$ as follows
\bes
\label{phih}
\begin{eqnarray}
\langle \Sigma \rangle=
\sigma \,
\left(
\begin{array}{cccc}
1&0&0&0 \\
0&1&0&0 \\
0&0&1&0 \\
0&0&0&-3 \\
\end{array}
\right) \,,
\end{eqnarray}
\begin{eqnarray}
\langle \Phi_H \rangle=
\left(
\begin{array}{cccc}
1&0&0&0 \\
0&1&0&0 \\
0&0&1&0 \\
0&0&0&-3 \\
\end{array}
\right) \otimes
\left(\begin{array}{cc}
v_H&0 \\
0&-v_H
\end{array}
\right)\
\,,
\end{eqnarray}
\ees
where the first matrix on the right-hand side of Eq. (\ref{phih})
refers to the direction $T_{15}$ of $SU(4)_{PS}$ whereas the
second matrix in the second equation refers to the direction $\tau_3$ 
in $SU(2)_H$, all of which refer to $(15,1,1,3)$. Here $\sigma$
and $v_H$ are constants.

When Eq. (\ref{yuk2}) is combined with Eq. (\ref{yuk1}), the
equation for the left-handed zero modes is now given by
\be
\label{zeroeq2}
\partial_{y}\xi_{L}(y)+\{f_{S}\, h_{S}(y)-f_{S'}\,\delta +
f_{H}\,\langle \Phi_H \rangle +f_{\Sigma}\langle \Sigma \rangle \}
\xi_{L}(y) = 0\,.
\ee

We now make an {\em important assumption}: 
\be
\label{equality}
f_H v_H = f_{\Sigma} \sigma \,.
\ee
This assumption has a far-reaching consequence: All unconventional
quarks and leptons will have large wave function overlaps
resulting in ``large'' mass scales for those sectors as
we shall see below.

Making use explicitely of Eq. (\ref{phih}) and the
assumption (\ref{equality}), one can now rewrite
(\ref{zeroeq2}) in terms various $SU(2)_L$ doublets as follows
\bes
\label{zeroeq3}
\be
\partial_{y}\xi_{L}^{Q}(y)+\{f_{S}\, h_{S}(y)+ 
2\,f_{H} v_H -f_{S'}\,\delta\}
\xi_{L}^{Q}(y) = 0\,,
\ee
\be
\partial_{y}\xi_{L}^{\tilde{Q}}(y)+\{f_{S}\, h_{S}(y)-
f_{S'}\,\delta\}
\xi_{L}^{\tilde{Q}}(y) = 0\,,
\ee
\be
\partial_{y}\xi_{L}^{L}(y)+\{f_{S}\, h_{S}(y)-6 
 f_{H} v_H-f_{S'}\,\delta\}
\xi_{L}^{L}(y) = 0\,,
\ee
\be
\partial_{y}\xi_{L}^{\tilde{L}}(y)+\{f_{S}\, h_{S}(y)
-f_{S'}\,\delta\}
\xi_{L}^{\tilde{L}}(y) = 0\,,
\ee
\ees
where the superscripts $Q,\tilde{Q},L,\tilde{L}$ refer to
the normal quark, unconventional quark, normal lepton,
and unconventional lepton $SU(2)_L$ doublets as shown in
Eq. (\ref{LR}). The above equations show the splitting between
conventional and unconventional fermions as well as between quarks 
and leptons. As we shall show below, this splitting is crucial
to the success of this model.

From Eqs. (\ref{zeroeq3}), it also is easy to see that
\be
\label{equall}
\xi_{L}^{\tilde{Q}}(y) = \xi_{L}^{\tilde{L}}(y) \,.
\ee

\subsubsection{The role of $\Phi_R = (1,1,3,1)$}
\label{phir}

We have encountered in Section \ref{LED} the $SU(2)_R$ triplet
scalar field whose kink solution, when combined with a singlet
kink, gives rise to very different profiles for the wave functions
of the up and down members of an $SU(2)_R$ fermion doublet. In the
present context, the triplet scalar field is now
$\Phi_R = (1,1,3,1)$. Its coupling to $\Psi^{\{R\}}$ can
be written as
\be
\label{yuk3}
{\cal L}_{Y3} = f_{R}\,
\bar{\Psi}^{\{R\}}\,\Phi_{R}\,\Psi^{\{R\}} \,,
\ee
with $f_R >0$.
Notice that here we use $f_R$ instead of the notation $f_T$ 
used in Section \ref{LED} in order to be consistent with
the notation used for $\Phi_{R}$.

The minimum energy solution for $\Phi_R$ can be written as
\begin{eqnarray}
\label{phir}
\langle \Phi_R \rangle=
\left(
\begin{array}{cccc}
1&0&0&0 \\
0&1&0&0 \\
0&0&1&0 \\
0&0&0&1 \\
\end{array}
\right) \otimes
\left(\begin{array}{cc}
h_{T}(y)&0 \\
0&-h_{T}(y)
\end{array}
\right)\
\,,
\end{eqnarray}
where we have assumed that there is a kink solution for $\Phi_R$.

When one combines Eq. (\ref{yuk1}) with Eqs. (\ref{yuk3},\ref{phir}), the
equation for the zero modes looks as follows.

\be
\label{zeroeq4}
\partial_{y}\xi_{R}(y)+\{f_{S}\, h_{S}(y)+ f_{R}\,\langle \Phi_R \rangle \}
\xi_{R}(y) = 0\,.
\ee
Let us define the following effective kinks:
\bes
\label{effkink}
\be
h^{sym}(y)=f_{S}\,h_{S}(y) + f_{R}\,h_{T}(y) \,,
\ee
\be
h^{asym}(y) = f_{S}\,h_{S}(y) - f_{R}\,h_{T}(y) \,.
\ee
\ees
Eq. (\ref{zeroeq4}) then takes the following forms:
\bes
\label{zeroeq5}
\be
\partial_{y}\xi_{R}^{Q,up}(y)+ h^{sym}(y)\,\xi_{R}^{Q,up}(y) = 0\,,
\ee
\be
\partial_{y}\xi_{R}^{Q,down}(y)+ h^{asym}(y)\,\xi_{R}^{Q,down}(y) = 0\,,
\ee
\be
\partial_{y}\xi_{R}^{L,up}(y)+ h^{sym}(y)\,\xi_{R}^{L,up}(y) = 0\,,
\ee
\be
\partial_{y}\xi_{R}^{L,down}(y)+ h^{asym}(y)\,\xi_{R}^{L,down}(y) = 0\,.
\ee
\ees

In Eqs. (\ref{zeroeq5}) and according to the particle content
given in Eqs. (\ref{LR}), ``$Q,up$'' refers to $d^{c}(1/3)$ 
and $\tilde{U}(4/3)$.
Likewise, ``$Q,down$'' refers to $u^{c}(-2/3)$ and $\tilde{D}(1/3)$.
Similarly, ``$L,up''$ refers to $\nu(0)$ and $\tilde{l}_{u}(-1)$ while
``$L,down''$ refers to $l(-1)$ and $\tilde{l}_{d}(-2)$. It is
also clear, from Eqs. (\ref{zeroeq5}), that
\bes
\label{equalr}
\be
\xi_{R}^{Q,up}(y) = \xi_{R}^{L,up}(y)=\xi_{R}^{up} \,,
\ee
\be
\xi_{R}^{Q,down}(y) = \xi_{R}^{L,down}(y)=\xi_{R}^{down} \,.
\ee
\ees

A quick look at Eq. (\ref{effkink}) reveals that 
$u_{R}(2/3)$ and $\tilde{D}_{R}(1/3)$ as well as
$l_{R}(-1)$ and $\tilde{l}_{d,R}(-2)$ have ``broad'' wave functions while
$d_{R}(-1/3)$ and $\tilde{U}_{R}(1/3)$ and
$\nu_{R}(0)$ and $\tilde{l}_{u,R}(-1)$ have ``narrow'' wave
functions. These features will be shown explicitly below.

From Eqs. (\ref{zeroeq5}), one can easily see that {\em all}
right-handed wave functions are localized at the origin. They have,
however, different profiles, a situation which is very similar
to the scenario which is summarized in Section \ref{LED}. It is
this difference in profiles, when combined with the different
locations of left-handed wave functions, which gives rise to
the disparity in mass scales.

We turn next our attention to the separations along $y$
between left-handed and right-handed fermions which are
crucial, along with the different profiles, in determining
the mass scale of each sector.

\subsubsection{Wave function localizations in a linear approximation}

To see heuristically how Eqs. (\ref{zeroeq3},\ref{zeroeq5}) 
help split the locations of the ``quarks'' and ``leptons''
along the extra dimension, let us make a linear approximation to the
kink solutions $h_{S}(y)$ and $h_{T}(y)$, namely 
\bes
\label{linear}
\be
h_{S}(y) \approx 2\, \mu_{S}^2 y \,.
\ee
\be
h_{T}(y) \approx 2\, \mu_{T}^2 y \,.
\ee
\ees

Let us recall that, with the linear approximation, a wave function behaves
like a Gaussian $\xi(y) \propto exp(-\mu^2 y^2)$. It then follows that
that the overlap between two functions separated by a distance $\Delta y$ along
$y$ goes like $exp(-\mu^2 (\Delta y)^2)$ (where for heuristic purpose $\mu^2$
are taken to be the same for both wave functions which in general is not the case).
The effective Yukawa couplings in four dimensions are proportional to the
overlaps between ``right-handed'' and ``left-handed'' fermions and it can 
be seen that they can be ``large'' or ``small'' depending
on whether or not $(\Delta y)^2 \gg \mu^2$ or $(\Delta y)^2 \ll \mu^2$.
 What we will
set out to derive in our model is the relationship between
$\Delta y^{(l)}$ and $\Delta y^{(q)}$. 

With the approximation (\ref{linear}), let us apply it to Eqs. (\ref{effkink}),
resulting in the following definitions
\bes
\label{mu}
\be
\mu_{asym}^{2} = f_{S}\,\mu_{S}^2 - f_{R}\,\mu_{T}^2 \,,
\ee
\be
\mu_{sym}^{2} =f_{S}\, \mu_{S}^2 + f_{R}\,\mu_{T}^2 \,,
\ee
\ees

From Eqs. (\ref{zeroeq5}), the locations of the Up-members and the
Down-members
of an $SU(2)_R$ doublet, for both conventional and unconventional 
quarks, are found to be at the {\em origin}. We shall denote that
common point by
\be
\label{locR}
y_R=0 \,.
\ee

For the left-handed zero modes, their locations will depend on
$f_{S'}\,\delta$ and $f_{H} v_{H}$. For convenience, let us
define the following quantity:
\be
\label{ratio}
r \equiv \frac{f_{S'}\,\delta}{2\,f_{H} v_{H}} \,.
\ee
We will assume that $r<1$.
From Eqs. (\ref{zeroeq3}), the locations of the $SU(2)_L$ doublets
for conventional and unconventional quarks are
\bes
\label{locQL}
\be
y_{Q_L} = (-\frac{f_{H}}{f_{S}})(\frac{v_H}{\mu_{S}^2})\,
(1-r) \,,
\ee
\be
y_{\tilde{Q}_L} = \frac{f_{S'}\,\delta}{2\,f_S \mu_{S}^2}=
(\frac{f_{H}}{f_{S}})(\frac{v_H}{\mu_{S}^2})\,r \,,
\ee
\ees
while the locations of the ``lepton'' doublets are
\bes
\label{loclL}
\be
y_{l_L} = (3\,\frac{f_{H}}{f_{S}})(\frac{v_H }{\mu_{S}^2})\,
(1+\frac{r}{3})\,,
\ee
\be
y_{\tilde{l}_L} = \frac{f_{S'}\,\delta}{2\,f_S \mu_{S}^2}=
(\frac{f_{H}}{f_{S}})(\frac{v_H}{\mu_{S}^2})\,r \,.
\ee
\ees

One important comment is in order at this point. From the above
equations (\ref{locR},\ref{locQL},\ref{loclL}) as
well as (\ref{mu}), it is clear that there are
{\em five independent parameters}: $f_{S}\,\mu_{S}^2$, $f_{S'}\,\delta$,
$f_{R}\,\mu_{T}^2$, $f_{\Sigma}\,\sigma$, and
$f_{H}\,v_H$, although Eq. (\ref{equality}) reduces to
{\em four independent parameters}. What this implies is that, 
out of {\em eight}
locations, there are {\em three} (or {\em four}) predictions. In principle,
we would then obtain {\em three} (or {\em four}) predictions for mass scales
once the other four (or three) are fixed by the choices of the aforementioned
parameters. We shall come back to this point below.

In order to make sense out of the above locations, a few remarks
are in order here. The effective Yukawa coupling, in four dimensions,
which governs the fermion mass scale depends on the overlap
between left-handed and right-handed fermions. This overlap depends on the separation between the two fermions as well as on the shapes
of the fermion wave functions. As mentioned above, the {\em spread}
of the wave functions is crucial in our scenario. This spread
is rougly {\em proportional} to $1/\mu$. From Eqs. (\ref{mu}), one
can deduce that, for the right-handed wave functions,
$l_{R}(-1)$ ,$\tilde{l}_{d,R}(-2)$, $u_{R}(2/3)$ and 
$\tilde{D}_{R}(1/3)$ have {\em broad} wave 
functions since $\mu_{asym}^{2} = f_{S}\,\mu_{S}^2 - \, f_{R}\,\mu_{T}^2$.
On the other hand, $d_{R}(-1/3)$, $\tilde{U}_{R}(4/3)$, 
$\nu_{R}(0)$ and $\tilde{l}_{u,R}(-1)$ have {\em narrow}
wave functions since
$\mu_{sym}^{2} =f_{S}\, \mu_{S}^2 + f_{R}\,\mu_{T}^2$. 
All the left-handed wave functions, on the other hand, have a spread
of the order of $1/\sqrt{f_{S}\,\mu_{S}^2}$.
How do these
facts translate into the disparities in mass scales? To answer this
question, one has to look at the separations between the left-handed
and right-handed wave functions.

From Eqs. (\ref{locR}-\ref{loclL}), one can readily derive the
following left-right separations.
\bi
\item For the ``quarks'':
\bes
\label{separationq}
\be
|\Delta y_{U}| \equiv |y_R - y_{Q_L}| = 
|(\frac{f_{H}}{f_{S}})(\frac{v_H}{\mu_{S}^2})\,
(1-r)| \,,
\ee
\be
|\Delta y_{D}| \equiv |y_R - y_{Q_L}| = 
|(\frac{f_{H}}{f_{S}})(\frac{v_H}{\mu_{S}^2})\,
(1-r)|\,,
\ee
\be
|\Delta y_{\tilde{U}}| \equiv |y_R - y_{\tilde{Q}_L}|=  
\frac{f_{S'}\,\delta}{2\,f_S \mu_{S}^2} =
(\frac{f_{H}}{f_{S}})(\frac{v_H}{\mu_{S}^2})\,r\,,
\ee
\be
|\Delta y_{\tilde{D}}| \equiv |y_R - y_{\tilde{Q}_L}| = 
\frac{f_{S'}\,\delta}{2\,f_S \mu_{S}^2}=
(\frac{f_{H}}{f_{S}})(\frac{v_H}{\mu_{S}^2})\,r \,,
\ee
\ees

\item For the ``leptons'':

\bes
\label{separationl}
\be
|\Delta y_{\nu}|\equiv |y_R - y_{l_L}|= 
3\,|(\frac{f_{H}}{f_{S}})(\frac{v_H }{\mu_{S}^2})\,
(1+\frac{r}{3})|\,,
\ee
\be
|\Delta y_{l(-1)}|\equiv |y_R - y_{l_L}|= 
3\,|(\frac{f_{H}}{f_{S}})(\frac{v_H }{\mu_{S}^2})\,
(1+\frac{r}{3})| \,,
\ee
\be
|\Delta y_{\tilde{l}_{u}(-1)}| \equiv |y_R - y_{\tilde{l}_L}|= 
\frac{f_{S'}\,\delta}{2\,f_S \mu_{S}^2}=
(\frac{f_{H}}{f_{S}})(\frac{v_H}{\mu_{S}^2})\,r\,,
\ee
\be
|\Delta y_{\tilde{l}_{d}(-2)}| \equiv |y_R - y_{\tilde{l}_L}| = 
\frac{f_{S'}\,\delta}{2\,f_S \mu_{S}^2}=
(\frac{f_{H}}{f_{S}})(\frac{v_H}{\mu_{S}^2})\,r\,.
\ee
\ees
\ei

Comparing Eqs. (\ref{separationl}) with Eqs. (\ref{separationq}), we
arrive at the following {\em important relationship} between {\em conventional}
quarks and leptons:
\be
\label{relation}
|\Delta y_{Lepton}| = 3\Biglb(\frac{1+\frac{r}{3}}{1-r}\Bigrb)\, 
|\Delta y_{Quark}| \,,
\ee
where $r<1$. For Scenario I, one would have $r=0$ and
one would simply have $|\Delta y_{Lepton}| = 3 |\Delta y_{Quark}|$.
For Scenario II, where $r \neq 0$, one obtains the above relationship.
What (\ref{relation}) implies is the following important fact:
$|\Delta y_{Lepton}| \geq 3 |\Delta y_{Quark}|$. This means that
the scales of the lepton sector are generally a bit smaller than those
of the quark counterpart.

It is also useful to derive a relationship between the separations
of conventional and unconventional fermions. From Eqs. (\ref{locQL},
\ref{loclL}, \ref{separationq}), it is straightforward to derive the
following relationship between the 
common left-right separation of the unconventional quarks and leptons 
and that of the conventional quarks:
\be
\label{relation2}
|\Delta y_{Unconventional}| = \Biglb(\frac{r}{1-r}\Bigrb)\, |\Delta y_{Quark}| \,.
\ee
Another useful form for (\ref{relation2}) can be obtained by using
Eq. (\ref{relation}), namely
\be
\label{relation3}
|\Delta y_{Unconventional}| = \Biglb(\frac{\frac{|\Delta y_{Lepton}|}
{|\Delta y_{Quark}|} -3}{4}\Bigrb) 
\, |\Delta y_{Quark}| \,.
\ee

From (\ref{relation}) and (\ref{relation2}, \ref{relation3}), one
notices that, in Scenario I where $r=0$, one obtains the simple
relations: $|\Delta y_{Lepton}| = 3 |\Delta y_{Quark}|$ and
$|\Delta y_{Unconventional}| = 0$.

The above relationship is important for the following reasons.
First, it implies that the left-handed wave function for the
leptons is situated much further (by a factor of three) away from
the right-handed ones than is the case for the quarks. It then
means that the wave function overlaps which
determine the effective four-dimensional Yukawa couplings would
, in principle, be much smaller for the lepton sector than for its
quark counterpart, resulting in a large disparity in mass scales
between the two sectors. This is actually what happens in reality.
The details of that disparity, in our scenario, will also depend
on the difference in the wave function profiles. This will be
discussed in the next section.

Since $|\Delta y|=\frac{f_{S'}\,\delta}{2\,f_S \mu_{S}^2}$ 
for both unconventional quarks and leptons, this implies that
unconventional quarks and leptons have comparable mass scales
which can be ``large''. How large this might be is the subject
of the next section. 

To summarize, the model contains 
{\em four independent parameters}: $f_{S}\,\mu_{S}^2$,
$f_{R}\,\mu_{T}^2$, $f_{H}\,v_H$, and
$r$. From these, we have two independent wave function
profiles for the right-handed zero modes, one independent
separation $|\Delta y_{Quark}|$, and one parameter $r$. Once
$r$ and $|\Delta y_{Quark}|$ are specified, all other separations
can be computed.

We now present some numerical illustrations of the above results. Our
strategy will be as follows. First, we write down the coupling
between the left-handed fermions, right-handed fermions and a Higgs
field whose VEV gives rise to fermion masses.
Next, we (arbitrarily) fix the two right-handed
wave function profiles. We then choose $|\Delta y_{Quark}|$ so
that the mass scales of the {\em conventional} Up and Down quark
sectors come out correctly. With the same wave function
profiles, we next choose $r$ so that, upon the use of Eq.(\ref{relation}),
the mass scale of the charged lepton sector comes out correctly.
We will show below that, as a result, we obtain {\em predictions}
for the mass scale of the {\em Dirac neutrino} sector as well
as those of the {\em unconventional} quarks and leptons. An
alternative way to fix the parameters is to choose one of
the right-handed wave functions, for example the one that belongs
to the charged leptons and the Up quark sector, fix $r$ so that the
mass scales come out correctly, fix the second right-handed 
wave function function so that the Down quark sector comes out
correctly. Once this is done, the above predictions will
come out the same.

\section{Computation of mass scales and implications}
\label{numeric}

One can now use the results of Ref. \cite{H2} and Section \ref{LED}
to estimate mass scales of the normal quark and lepton sectors
as well as those of the unconventional fermions. Since the
scope of this paper is the construction of a model showing
a relationship between mass scales of ``quarks'' and ``lepton''
sectors, we shall ignore issues such as fermion mixings
in the mass matrices. Higher dimensional models have been
built to tackle quark mass hierarchies, mixing angles and CP phase
(see e.g. \cite{HS}and \cite{HSS}). We will therefore 
concentrate on the overall mass
scales that appear in various mass matrices.

\subsection{SM Fermion-Higgs coupling}
\label{SMhiggs}

By ``SM Fermion-Higgs coupling'', we mean that the Higgs
field that couples with left-handed and right-handed fermions
transforms non-trivially under $SU(2)_L$.

Since $\Psi^{\{L\}}= (4,2,1,2)$, $\Psi^{\{R\}} = (4,1,2,2)$
and $\bar{\Psi^{\{L\}}}\,\Psi^{\{R\}} = (1+15,2,2,1+3)$, an
appropriate Higgs field (the simplest choice) could be the
following field:
\be
\label{higgs}
H = (1,2,2,1) \,.
\ee
The Yukawa coupling with this field can be written as
\be
\label{yuk4}
{\cal L}_{Y4} = k_{1}\bar{\Psi}^{\{L\}}H\,\Psi^{\{R\}}+
k_{2}\bar{\Psi}^{\{L\}}\tilde{H}\,\Psi^{\{R\}}+H.c. \,,
\ee
where $\tilde{H} = \tau_{2} H^{\ast} \tau_{2}$
and where, in principle, $k_{2}$ {\em can be different}
from $k_{1}$.
Assuming the extra dimension to be compactified on an orbifold
$S_1 / Z_2$ and an even $Z_2$ parity for $H$, it follows that
$H$ can have a {\em zero mode}. This zero mode can be written
as $H^{0}(x,y) = K \phi(x)$ where $\phi(x)$ is a
4-dimensional Higgs field with dimension $M$ (mass) 
and $K$, a constant,
has a dimension $\sqrt{M}$, since $H$ has a dimension $M^{3/2}$
in five dimensions. Notice that $k_{1,2}$ have a dimension
$M^{-1/2}$. We define the following dimensionless couplings:
\be
\label{dless}
g_{Y1,2} = k_{1,2}K \,.
\ee

The VEV of $\phi$ is assumed to be
\be
\label{phi}
\langle \phi \rangle =
\left(\begin{array}{cc}
v_{1}/\sqrt{2}&0 \\
0&v_{2}/\sqrt{2}
\end{array}
\right)\  \,.
\ee
Eqs. (\ref{yuk4}) and (\ref{phi}) provide mass scales
which appear in mass matrices as follows
\be
\label{massmatrix}
{\cal M}_{U,D,\nu,l^{-}, \tilde{U}, \tilde{D},
\tilde{l}_u, \tilde{l}_d} =
\Lambda_{U,D,\nu,l^{-}, \tilde{U}, \tilde{D},
\tilde{l}_u, \tilde{l}_d}\,
M_{U,D,\nu,l^{-}, \tilde{U}, \tilde{D},
\tilde{l}_u, \tilde{l}_d} \,,
\ee
where $\Lambda_{U,D,\nu,l^{-}, \tilde{U}, \tilde{D},
\tilde{l}_u, \tilde{l}_d}$ are the mass scales in question and
$M_{U,D,\nu,l^{-}, \tilde{U}, \tilde{D},
\tilde{l}_u, \tilde{l}_d}$ are matrices whose elements will
depend on models of fermion masses (e.g. \cite{HS}).
The subscripts are self-explanatory.

Using the fermion representations (\ref{LR}) and
Eqs. (\ref{yuk4}, \ref{phi}, \ref{equall}, \ref{equalr}), 
the mass scales that
appear in Eq. (\ref{massmatrix}) now take the following
forms:
\bes
\label{scales}
\be
\Lambda_{U} = g_{U1}\,v_2/\sqrt{2} + g_{U2}\,v_1/\sqrt{2} \,,
\ee
\be
\Lambda_{D} = g_{D1}\,v_1/\sqrt{2} + g_{D2}\,v_2/\sqrt{2} \,,
\ee
\be
\Lambda_{\nu} = g_{\nu 1}\,v_1/\sqrt{2} + g_{\nu 2}\,v_2/\sqrt{2} \,,
\ee
\be
\Lambda_{l^{-}} = g_{l^{-} 1}\,v_2/\sqrt{2} + g_{l^{-} 2}\,v_1/\sqrt{2} \,,
\ee
\be
\Lambda_{\tilde{U}} = \Lambda_{\tilde{l}_u} = 
g_{\tilde{U} 1}\,v_1/\sqrt{2} + g_{\tilde{U} 2}\,v_2/\sqrt{2} \,,
\ee
\be
\Lambda_{\tilde{D}} = \Lambda_{\tilde{l}_d} = 
g_{\tilde{D} 1}\,v_2/\sqrt{2} + g_{\tilde{D} 2}\,v_1/\sqrt{2} \,,
\ee
\ees 
where
\bes
\label{effyuk}
\be
g_{U1,2} = g_{Y1,2}\,\int_{0}^{L} dy \,\xi_{L}^{Q}(y)\,
\xi_{R}^{down}(y)  \,,
\ee
\be
g_{D1,2} = g_{Y1,2}\,\int_{0}^{L} dy \,\xi_{L}^{Q}(y)\,
\xi_{R}^{up}(y)  \,,
\ee
\be
g_{\nu 1,2} = g_{Y1,2}\,\int_{0}^{L} dy \,\xi_{L}^{L}(y)\,
\xi_{R}^{up}(y) \,,
\ee
\be
g_{l^{-} 1,2} = g_{Y1,2}\,\int_{0}^{L} dy \,\xi_{L}^{L}(y)\,
\xi_{R}^{down}(y) \,,
\ee
\be
g_{\tilde{U}1,2} =g_{\tilde{l}_{u} 1,2} =
g_{Y1,2}\,\int_{0}^{L} dy \,\xi_{L}^{\tilde{Q}}(y)\,
\xi_{R}^{up}(y)  \,,
\ee
\be
g_{\tilde{D}1,2} = g_{\tilde{l}_{d} 1,2}=
g_{Y1,2}\,\int_{0}^{L} dy \,\xi_{L}^{\tilde{Q}}(y)\,
\xi_{R}^{down}(y)  \,.
\ee
\ees
The way $v_1$ and $v_2$ appear in Eqs. (\ref{scales}) should be 
clearly understood that it has to do with the fermion content as
shown in Eqs. (\ref{LR}), and that is the reason why the first
two equations appear in the forms shown above.


There are two possibilities concerning Eqs. (\ref{scales}).

\bi

\item $g_{Y1} = g_{Y2}$:

This is a rather economical option, in terms of reducing the
number of parameters.
From Eqs. (\ref{scales}) and
(\ref{effyuk}), it is easy to see that,
{\em if} $g_{Y1} = g_{Y2}$, the ratios of scales will
simply {\em ratios of wave function overlaps} regardless
of the values of $v_1$ and $v_2$ as well as of the value
of the Yukawa coupling since they cancel out in the ratios.
In fact, we can form {\em six} ratios (from six independent
scales) which are can be taken as $\Lambda_{D}/
\Lambda_{U}$, $\Lambda_{\nu}/\Lambda_{l^{-}}$, 
$\Lambda_{\nu}/\Lambda_{D}$,
$\Lambda_{l^{-}}/\Lambda_{U}$, $\Lambda_{\tilde{U}}/
\Lambda_{D}$, $\Lambda_{\tilde{D}}/\Lambda_{U}$. 
Explicitely, one has:
\bes
\label{ratios}
\be
\frac{\Lambda_{D}}{\Lambda_{U}} = \frac{
\int_{0}^{L} dy \,\xi_{L}^{Q}(y)\,
\xi_{R}^{up}(y)}{
\int_{0}^{L} dy \,\xi_{L}^{Q}(y)\,
\xi_{R}^{down}(y)} \,,
\ee
\be
\frac{\Lambda_{\nu}}{\Lambda_{l^{-}}} = \frac{
\int_{0}^{L} dy \,\xi_{L}^{L}(y)\,
\xi_{R}^{up}(y)}{
\int_{0}^{L} dy \,\xi_{L}^{L}(y)\,
\xi_{R}^{down}(y)} \,,
\ee
\be
\frac{\Lambda_{\nu}}{\Lambda_{D}} = \frac{
\int_{0}^{L} dy \,\xi_{L}^{L}(y)\,
\xi_{R}^{up}(y)}{
\int_{0}^{L} dy \,\xi_{L}^{Q}(y)\,
\xi_{R}^{up}(y)} \,,
\ee
\be
\frac{\Lambda_{l^{-}}}{\Lambda_{U}} = \frac{
\int_{0}^{L} dy \,\xi_{L}^{L}(y)\,
\xi_{R}^{down}(y)}{
\int_{0}^{L} dy \,\xi_{L}^{Q}(y)\,
\xi_{R}^{down}(y)} \,,
\ee
\be
\frac{\Lambda_{\tilde{U}}}{\Lambda_{D}} = \frac{
\int_{0}^{L} dy \,\xi_{L}^{\tilde{Q}}(y)\,
\xi_{R}^{up}(y)}{
\int_{0}^{L} dy \,\xi_{L}^{Q}(y)\,
\xi_{R}^{up}(y)} \,,
\ee
\be
\frac{\Lambda_{\tilde{D}}}{\Lambda_{U}} = \frac{
\int_{0}^{L} dy \,\xi_{L}^{\tilde{Q}}(y)\,
\xi_{R}^{down}(y)}{
\int_{0}^{L} dy \,\xi_{L}^{Q}(y)\,
\xi_{R}^{down}(y)} \,,
\ee
\ees

Once
the parameters of the wave functions and their separations are
fixed, these ratios (or any other combinations) can be computed
unambigously.

In the following,
we will choose $\Lambda_{U}$ and $\Lambda_{D}$ as two
independent inputs. From them, we can extract $|\Delta y_{Quark}|$.
Once the parameter $r$ is chosen, all other mass scales can
be predicted.

\item $g_{Y1} \neq g_{Y2}$:

For the case when $g_{Y1} \neq g_{Y2}$,
one can still obtain the following ratios which
depend only on ratios of wave function overlaps: 
\bes
\label{ratios2}
\be
\frac{\Lambda_{l^{-}}}{\Lambda_{U}} = \frac{
\int_{0}^{L} dy \,\xi_{L}^{L}(y)\,
\xi_{R}^{down}(y)}{
\int_{0}^{L} dy \,\xi_{L}^{Q}(y)\,
\xi_{R}^{down}(y)} \,,
\ee
\be
\frac{\Lambda_{\nu}}{\Lambda_{D}} = \frac{
\int_{0}^{L} dy \,\xi_{L}^{L}(y)\,
\xi_{R}^{up}(y)}{
\int_{0}^{L} dy \,\xi_{L}^{Q}(y)\,
\xi_{R}^{up}(y)} \,,
\ee
\be
\frac{\Lambda_{\tilde{U}}}{\Lambda_{D}} = \frac{
\int_{0}^{L} dy \,\xi_{L}^{\tilde{Q}}(y)\,
\xi_{R}^{up}(y)}{
\int_{0}^{L} dy \,\xi_{L}^{Q}(y)\,
\xi_{R}^{up}(y)} \,,
\ee
\be
\frac{\Lambda_{\tilde{D}}}{\Lambda_{U}} = \frac{
\int_{0}^{L} dy \,\xi_{L}^{\tilde{Q}}(y)\,
\xi_{R}^{down}(y)}{
\int_{0}^{L} dy \,\xi_{L}^{Q}(y)\,
\xi_{R}^{down}(y)} \,,
\ee
\ees

\ei

What are the implications of the above two cases? First, one chooses the 
two parameters $f_{S}\,\mu_{S}^2$ and $f_{R}\,\mu_{T}^2$ so that 
$\xi_{R}^{up}(y)$ and $\xi_{R}^{down}(y)$ are fixed. Next, we
choose $\Lambda_{U}$ and $\Lambda_{D}$ as two independent mass
scales. In the first case where $g_{Y1} = g_{Y2}$, these two scales
are used to extract $|\Delta y_{Quark}|$. One can then choose
the parameter $r$ so that the scale $\Lambda_{l^{-}}$ is fixed. Once
this is done, {\em all} other scales- neutrinos, unconventional fermions-
can be predicted. For the second case where $g_{Y1} \neq g_{Y2}$, one
has to {\em both} choose $|\Delta y_{Quark}|$ and $r$ in order to fix
the first ratio in (\ref{ratios2}). All other ratios in (\ref{ratios2})
can then be predicted.


\subsection{Some numerical examples}

In this section, we will present some numerical illustrations of the
above ideas. A more comprehensive numerical analysis will be presented 
elsewhere. We find a surprising correlation between the Dirac
neutrino masses and those of the unconventional fermions. As
we shall see below, by requiring the unconventional fermions
to be heavier than the top quark but at the same time NOT too
much heavier than the electroweak scale e.g. 
$< 650\,GeV$, it is found that
the largest {\em Dirac} neutrino mass is bounded from below by a
value roughly of the order of $0.1 \, eV$ and from above by
a value of the order of $30 \, eV$. From this result alone,
it is hard to see how one can incorporate Majorana neutrinos
in our scenario since the Dirac neutrinos {\em alone} are naturally
light. Actually, the detailed numerical analysis of
(\cite{BH2}) shows that, in order to keep the early unification
scale below $2\, TeV$, the masses of the unconventional fermions
are constrained to be less than $300\,TeV$ which, as we shall
see below, sets the following bound for the heaviest
Dirac neutrino: $0.1 \, eV < m_{\nu}^{heaviest} < 1 \,eV$.
Or one can turn things around by using some of our
knowledge, however uncertain. about neutrino masses to
set bounds on the masses of the unconventional fermions.
For example, if we set the largest neutrino mass to
lie between $0.1\,eV$ and $1\,eV$ assuming it is of the
Dirac type, the unconventional fermions are constrained
to have a mass between the top quark mass and $300\,GeV$.

Our numerical strategy is as follows. 1) For a given pair of
$f_{S}\,h_{S}(y)$ and $f_{R}\,h_{T}(y)$, we use the ratio
$\Lambda_{D}/\Lambda_{U}$ to find $|\Delta y_{Quark}|$. 
Actually, it is the difference between $f_{S}\,h_{S}(y)$ and 
$f_{R}\,h_{T}(y)$ that is important. 2) We
then choose $r$ so that $|\Delta y_{Lepton}|$, which is given
in terms of $|\Delta y_{Quark}|$ via Eq. (\ref{relation}),
gives the correct ratio $\Lambda_{l^{-}}/\Lambda_{U}$. 3)
After Steps 1 and 2 have been carried out, one can make
{\em predictions} for the mass scales of the {\em Dirac} 
neutrino sector
as well as those for the unconventional fermions, using
Eqs. (\ref{ratios}). After this is done, one can then
decide whether or not a Majorana neutrino mass term is needed,
depending on the value of the Dirac mass.

To be general, we start out with $r \neq 0$.

For the zero mode right-handed wave functions, we use expressions 
similar to those found in Eq. (\ref{xinu2}), namely
\begin{equation}
\label{xinu3}
\xi_{R}^{up, down}(y) = k_{up, down} e^{-(C_{S} \ln(\cosh(\mu_{S}y)) \pm
C_{R} \ln(\cosh(\mu_{T}y)))} \, ,
\end{equation}
where $k_{up, down}$ are normalization factors and
$C_{S,R} = f_{S,R}/(\lambda_{S,T}/2)^{1/2}$ are factors which involve
the Yukawa couplings as well as the self-couplings in the scalar
potentials. (Let us recall generically that $h(y)=v\,\tanh (\mu y)$, with 
$\mu =\sqrt{\lambda/2}$.) The wave functions for the left-handed
zero modes are given by
\begin{equation}
\label{xinu4}
\xi_{L}^{i}(y) = k_{L} e^{-(C_{S} \ln(\cosh(\mu_{S}(y-y_{i}))))} \, ,
\end{equation}
where $k_L$ is a normalization factor, $i = Q, L, \tilde{Q}$ and
$y_i$ are given by (\ref{locQL}) and (\ref{loclL}).

Using (\ref{xinu3}) and (\ref{xinu4}), we can now illustrate the
results presented in the previous section with a numerical
example. 
To translate ratios of mass scales into ratios of actual mass
eigenvalues, an assumption has to be made concerning the mass
matrices themselves. Explicitely, the relationship
between the mass scales $\Lambda$ and the mass matrices 
${\cal M}$ can be written as
\be
\label{massmatrix}
{\cal M} = \Lambda \, \bar{{\cal M}} \,,
\ee
where $\bar{{\cal M}}$ is a dimensionless matrix whose form
depends on a particular model.
An exhaustive general analysis of different types
of mass matrices is beyond the scope
of this paper. For the purpose of illustration in this
paper, we will make the following reasonable assumptions
concerning the relationship between the mass scale that
appears as a common factor in the mass matrix and the
largest eigenvalue, namely
\bes
\label{scale}
\be
\frac{m_t}{3} \leq \Lambda_{U} \leq m_t \,,
\ee
\be
\frac{m_b}{3} \leq \Lambda_{D} \leq m_b \,,
\ee
\be
\frac{m_{\tau}}{3} \leq \Lambda_{l^{-}} \leq m_{\tau} \,,
\ee
\ees
where $m_t$, $m_b$, and $m_{\tau}$ are the largest eigenvalues
of the Up-quark, Down-quark, and charged lepton mass matrices
respectively. The lower bounds in Eqs. (\ref{scale}) refer to
a pure democratic mass matrix \cite{democratic} 
where, apart from the common 
scale factor, all elements are unity and the largest eigenvalue
is {\em three times} the scale factor. Such a pure democratic mass matrix
is unrealistic but it is included here for completeness. The upper
bound refers to a class of hierarchical mass matrices where the 
largest eigenvalue is approximately the scale factor itself
\cite{hierarchical}.
In between these two bounds, there exists models e.g. (\cite{HS,
HSS}) which are almost but not quite of the pure phase mass matrix
type \cite{PPM}. We will assume below that the Up-quark, Down-quark 
and charged lepton sectors have mass matrices with ``similar''
behaviour, only in the sense that the relationship between the
scale factors and the largest mass eigenvalues is assumed to
be the same. 

For the purpose of illustration, we will use, for the largest
eigenvalues, $m_t$, $m_b$ and $m_{\tau}$ evaluated at $M_Z$,
and neglect any ``running'' between $M_Z$ and the early
unification scale taken to be comparable to the scale of
compactification. We take
\be
\label{running}
m_t(M_Z) = 181\, GeV; \,m_b(M_Z) = 3\, GeV;\, m_{\tau}(M_Z) = 1.747\,GeV\,. 
\ee
With the above numbers and with the remarks made above concerning
the relationship between the scale factors and the largest mass 
eigenvalues, we can write
\bes
\be
\label{numericsq}
\frac{\Lambda_{D}}{\Lambda_{U}} \approx \frac{m_b(M_Z)}
{m_t(M_Z)} \approx 0.0166 \,,
\ee
\be
\label{numericsl}
\frac{\Lambda_{l^{-}}}{\Lambda_{U}} \approx \frac{m_{\tau}(M_Z)}
{m_t(M_Z)} \approx 0.00965 \,.
\ee
\ees
The ratios (\ref{numericsq},\ref{numericsl}) are now used 
to {\em predict} the mass
scales $\Lambda$ for the neutrino sector as well as for the 
unconventional fermion sectors.

As we have mentioned earlier, the mass scales of the neutrinos
are correlated with those of the unconventional fermions. This
will be shown in the six examples below for the more general
case of $r \neq 0$. For comparison, we will also show a result
for $r = 0$.

First we would like to remind ourselves that it is the
difference between $C_{S} \ln(\cosh(\mu_{S}y))$ and
$C_{R} \ln(\cosh(\mu_{T}y))$ in the wave functions 
that is important. In consequence, we will set $C_{S}=
C_{R} =1$, choose $\mu_{S} =1$ (in some unit), and
vary $\mu_{T}$.

In order to present the results in a transparent way, we
prefer to write expressions such as $\exp (-\ln(\cosh(y)))$
instead of the equivalent expression $1/\cosh(y)$.

\bi

\item  $\xi_{R}^{up}(y)= \frac{1}{\sqrt{1.553}}\,\exp\{-\ln(\cosh(y)) -
\ln(\cosh(0.7\,y))\}$; 
$\xi_{R}^{down}(y)= \frac{1}{\sqrt{3.718}}\,\exp\{-\ln(\cosh(y)) +
\ln(\cosh(0.7\,y))\}$;
$\xi_{L}^{Q}(y)=\frac{1}{\sqrt{2}}\,\exp\{-\ln(\cosh(y+7.815))\}$;
$\xi_{L}^{L}(y)=\frac{1}{\sqrt{2}}\,\exp\{-\ln(\cosh(y-23.285))\}$;
$\xi_{L}^{\tilde{Q}}(y)=\frac{1}{\sqrt{2}}\,\exp\{-\ln(\cosh(y+0.04))\}$.

Here $y_{Q} = -7.815$ is chosen for a given pair $\xi_{R}^{up,down}(y)$
so that the ratio (\ref{numericsq}) is satisfied. Similarly,
$y_{L}=23.285$ is chosen so that the ratio (\ref{numericsl}) is satisfied.
The location $y_{\tilde{Q}} = -0.04$ for the unconventional fermions
was calculated using Eq. (\ref{relation3}). The predictions for the
neutrino and unconventional fermion mass scales are found to be
\bes
\label{rat1}
\be
\frac{\Lambda_{\nu}}{\Lambda_{l^{-}}} =0.278 \times 10^{-6} \,,
\ee
\be
\frac{\Lambda_{\tilde{U}}}{\Lambda_{U}} =
\frac{\Lambda_{\tilde{l}_u}}{\Lambda_{U}}= 7.3 \,,
\ee
\be
\frac{\Lambda_{\tilde{D}}}{\Lambda_{U}} =
\frac{\Lambda_{\tilde{l}_d}}{\Lambda_{U}}= 7.93 \,.
\ee
\ees

\item $\xi_{R}^{up}(y)= \frac{1}{\sqrt{1.53}}\,\exp\{-\ln(\cosh(y)) -
\ln(\cosh(0.73\,y))\}$; 
$\xi_{R}^{down}(y)= \frac{1}{\sqrt{4.057}}\,\exp\{-\ln(\cosh(y)) +
\ln(\cosh(0.73\,y))\}$;
$\xi_{L}^{Q}(y)=\frac{1}{\sqrt{2}}\,\exp\{-\ln(\cosh(y+7.53))\}$;
$\xi_{L}^{L}(y)=\frac{1}{\sqrt{2}}\,\exp\{-\ln(\cosh(y-24.715))\}$;
$\xi_{L}^{\tilde{Q}}(y)=\frac{1}{\sqrt{2}}\,\exp\{-\ln(\cosh(y-0.531))\}$.

The predictions are:
\bes
\label{rat2}
\be
\frac{\Lambda_{\nu}}{\Lambda_{l^{-}}} = 0.5 \times 10^{-7} \,,
\ee
\be
\frac{\Lambda_{\tilde{U}}}{\Lambda_{U}} =
\frac{\Lambda_{\tilde{l}_u}}{\Lambda_{U}}= 5.46 \,,
\ee
\be
\frac{\Lambda_{\tilde{D}}}{\Lambda_{U}} =
\frac{\Lambda_{\tilde{l}_d}}{\Lambda_{U}}= 5.82 \,.
\ee
\ees

\item $\xi_{R}^{up}(y)= \frac{1}{\sqrt{1.514}}\,\exp\{-\ln(\cosh(y)) -
\ln(\cosh(0.75\,y))\}$; 
$\xi_{R}^{down}(y)= \frac{1}{\sqrt{4.332}}\,\exp\{-\ln(\cosh(y)) +
\ln(\cosh(0.75\,y))\}$;
$\xi_{L}^{Q}(y)=\frac{1}{\sqrt{2}}\,\exp\{-\ln(\cosh(y+7.53))\}$;
$\xi_{L}^{L}(y)=\frac{1}{\sqrt{2}}\,\exp\{-\ln(\cosh(y-24.715))\}$;
$\xi_{L}^{\tilde{Q}}(y)=\frac{1}{\sqrt{2}}\,\exp\{-\ln(\cosh(y-0.531))\}$.

The predictions are:
\bes
\label{rat3}
\be
\frac{\Lambda_{\nu}}{\Lambda_{l^{-}}} = 0.13 \times 10^{-7} \,,
\ee
\be
\frac{\Lambda_{\tilde{U}}}{\Lambda_{U}} =
\frac{\Lambda_{\tilde{l}_u}}{\Lambda_{U}}= 4.43 \,,
\ee
\be
\frac{\Lambda_{\tilde{D}}}{\Lambda_{U}} =
\frac{\Lambda_{\tilde{l}_d}}{\Lambda_{U}}= 4.37 \,.
\ee
\ees

\item $\xi_{R}^{up}(y)= \frac{1}{\sqrt{1.483}}\,\exp\{-\ln(\cosh(y)) -
\ln(\cosh(0.79\,y))\}$; 
$\xi_{R}^{down}(y)= \frac{1}{\sqrt{5.049}}\,\exp\{-\ln(\cosh(y)) +
\ln(\cosh(0.79\,y))\}$;
$\xi_{L}^{Q}(y)=\frac{1}{\sqrt{2}}\,\exp\{-\ln(\cosh(y+7.07))\}$;
$\xi_{L}^{L}(y)=\frac{1}{\sqrt{2}}\,\exp\{-\ln(\cosh(y-29.17))\}$;
$\xi_{L}^{\tilde{Q}}(y)=\frac{1}{\sqrt{2}}\,\exp\{-\ln(\cosh(y-1.99))\}$.

The predictions are:
\bes
\label{rat4}
\be
\frac{\Lambda_{\nu}}{\Lambda_{l^{-}}} = 0.381 \times 10^{-9} \,,
\ee
\be
\frac{\Lambda_{\tilde{U}}}{\Lambda_{U}} =
\frac{\Lambda_{\tilde{l}_u}}{\Lambda_{U}}= 2.834 \,,
\ee
\be
\frac{\Lambda_{\tilde{D}}}{\Lambda_{U}} =
\frac{\Lambda_{\tilde{l}_d}}{\Lambda_{U}}= 1.858 \,.
\ee
\ees

\item $\xi_{R}^{up}(y)= \frac{1}{\sqrt{1.476}}\,\exp\{-\ln(\cosh(y)) -
\ln(\cosh(0.8\,y))\}$; 
$\xi_{R}^{down}(y)= \frac{1}{\sqrt{5.276}}\,\exp\{-\ln(\cosh(y)) +
\ln(\cosh(0.8\,y))\}$;
$\xi_{L}^{Q}(y)=\frac{1}{\sqrt{2}}\,\exp\{-\ln(\cosh(y+7))\}$;
$\xi_{L}^{L}(y)=\frac{1}{\sqrt{2}}\,\exp\{-\ln(\cosh(y-30.2))\}$;
$\xi_{L}^{\tilde{Q}}(y)=\frac{1}{\sqrt{2}}\,\exp\{-\ln(\cosh(y-2.3))\}$.

The predictions are:
\bes
\label{rat5}
\be
\frac{\Lambda_{\nu}}{\Lambda_{l^{-}}} = 0.13 \times 10^{-9} \,,
\ee
\be
\frac{\Lambda_{\tilde{U}}}{\Lambda_{U}} =
\frac{\Lambda_{\tilde{l}_u}}{\Lambda_{U}}= 2.521 \,,
\ee
\be
\frac{\Lambda_{\tilde{D}}}{\Lambda_{U}} =
\frac{\Lambda_{\tilde{l}_d}}{\Lambda_{U}}= 1.39 \,.
\ee
\ees

\item $\xi_{R}^{up}(y)= \frac{1}{\sqrt{1.468}}\,\exp\{-\ln(\cosh(y)) -
\ln(\cosh(0.81\,y))\}$; 
$\xi_{R}^{down}(y)= \frac{1}{\sqrt{5.527}}\,\exp\{-\ln(\cosh(y)) +
\ln(\cosh(0.81\,y))\}$;
$\xi_{L}^{Q}(y)=\frac{1}{\sqrt{2}}\,\exp\{-\ln(\cosh(y+6.94))\}$;
$\xi_{L}^{L}(y)=\frac{1}{\sqrt{2}}\,\exp\{-\ln(\cosh(y-31.36))\}$;
$\xi_{L}^{\tilde{Q}}(y)=\frac{1}{\sqrt{2}}\,\exp\{-\ln(\cosh(y-2.635))\}$.

The predictions are:
\bes
\label{rat6}
\be
\frac{\Lambda_{\nu}}{\Lambda_{l^{-}}} = 0.371 \times 10^{-10} \,,
\ee
\be
\frac{\Lambda_{\tilde{U}}}{\Lambda_{U}} =
\frac{\Lambda_{\tilde{l}_u}}{\Lambda_{U}}= 2.246 \,,
\ee
\be
\frac{\Lambda_{\tilde{D}}}{\Lambda_{U}} =
\frac{\Lambda_{\tilde{l}_d}}{\Lambda_{U}}= 0.998 \,.
\ee
\ees

\ei

The implications of the above results are given for
the upper and lower bounds in (\ref{scale}) 
in Tables I and II.
We use the values in Eq. (\ref{running}) for the estimates given
in these tables. The predictions coming from (\ref{rat1}-\ref{rat6}) 
are listed in the second, third, and fourth columns.

\begin{table}
\caption{\label{tab:table1}. Predictions for
$\Lambda_{\nu}$, $\Lambda_{\tilde{U}}=\Lambda_{\tilde{l}_u}$,
and $\Lambda_{\tilde{D}}=\Lambda_{\tilde{l}_d}$ for the
upper bounds (\ref{scale}):
$\Lambda_{U} \approx m_t(M_Z)$, $\Lambda_{D} 
\approx m_b(M_Z)$,and $\Lambda_{l^{-}} \approx m_{\tau}(M_Z)$}
\begin{ruledtabular}
\begin{tabular}{ccccc} 
&$\Lambda_{\nu} \approx$&$\Lambda_{\tilde{U}} \approx$
&$\Lambda_{\tilde{D}} \approx$ \\ \hline
Eq. (\ref{rat1})&$486\,eV$&$1321\,GeV$&$1435\,GeV$ \\
Eq. (\ref{rat2})&$87 \,eV$&$988\,GeV$&$1053\,GeV$ \\
Eq. (\ref{rat3})&$23\,eV$&$802\,GeV$&$791\,GeV$ \\
Eq. (\ref{rat4})&$0.67 \,eV$&$513\,GeV$&$336\,GeV$ \\
Eq. (\ref{rat5})&$0.23\,eV$&$456\,GeV$&$252\,GeV$ \\
Eq. (\ref{rat6})&$0.065\,eV$&$406\,GeV$&$181\,GeV$ \\
\end{tabular}
\end{ruledtabular}
\end{table}

\begin{table}
\caption{\label{tab:table2}. Predictions for
$\Lambda_{\nu}$, $\Lambda_{\tilde{U}}=\Lambda_{\tilde{l}_u}$,
and $\Lambda_{\tilde{D}}=\Lambda_{\tilde{l}_d}$ for the
lower bounds (\ref{scale}):
$\Lambda_{U} \approx m_t(M_Z)/3$, $\Lambda_{D} 
\approx m_b(M_Z)/3$,and $\Lambda_{l^{-}} \approx m_{\tau}(M_Z)/3$}
\begin{ruledtabular}
\begin{tabular}{ccccc} 
&$\Lambda_{\nu} \approx$&$\Lambda_{\tilde{U}} \approx$
&$\Lambda_{\tilde{D}} \approx$ \\ \hline
Eq. (\ref{rat1})&$162\,eV$&$440\,GeV$&$478\,GeV$ \\
Eq. (\ref{rat2})&$29\,eV$&$329\,GeV$&$351\,GeV$ \\
Eq. (\ref{rat3})&$7.7\,eV$&$267\,GeV$&$264\,GeV$ \\
Eq. (\ref{rat4})&$0.22 \,eV$&$171\,GeV$&$112\,GeV$ \\
Eq. (\ref{rat5})&$0.077\,eV$&$152\,GeV$&$84\,GeV$ \\
Eq. (\ref{rat6})&$0.022\,eV$&$135\,GeV$&$60\,GeV$ \\
\end{tabular}
\end{ruledtabular}
\end{table}

We end this section by briefly showing that the case $r=0$ which gives
the interesting relations 
$|\Delta y_{Lepton}| = 3 |\Delta y_{Quark}|$ and
$|\Delta y_{Unconventional}| = 0$ and which means that one can also
predict the charged lepton mass scale in terms of the one for the quarks,
is, unfortunately, not good. For example, taking the quark wave functions
used in (\ref{rat5}) and using the above relations, one obtains
a prediction for the mass scale of the charged lepton sector to
be approximately $11\,GeV$. This, by itself, rules out the case
$r=0$. Incidentally, the neutrino mass scale comes out to be
$\sim 2.5\,keV$ and those of the unconventional fermions
come out to be $\sim 590-700\,GeV$, although these numbers are 
irrelevant since the prediction for the charged lepton sector is already
wrong.

We now discuss the implications of Tables I and II for the more
general case $r \neq 0$.

\subsection{Implications of Tables I and II}

To obtain a better understanding of the numerical results presented
in Tables I and II, we will assume that the mass matrices for
the unconventional fermions are such that, for each sector, the fermions
are approximately degenerate. That is because, if it were not the
case, a mass splitting similar to the normal fermions (quarks and
charged leptons) would renders at least one fermion for each sector
to be lighter that say the top quark. It goes without saying that
none has been seen so far. We will therefore assume that the masses
of the unconventional fermions for each sector are approximately
equal to the mass scales $\Lambda$'s.

\bi

\item Table I:

The numerical results given in this table are for the case where
the mass matrices of the normal quarks and leptons are highly
hierarchical as we had mentioned earlier.

One obvious remark that one can make by looking at Table I is the following.
There is a clear relationship between the masses of the unconventional
fermions and those of the neutrinos: As the unconventional masses
increase so do neutrino masses. However,
if we restrict the masses of the unconventional fermions to be {\em less}
than one TeV, one notices that the {\em Dirac} mass of the neutrinos
cannot exceed a few hundreds $eV$s. In scenarios such as this one,
one might expect that Majorana masses, if they exist, would typically
be also of the order of TeVs. The see-saw mass for the light neutrino
would then be roughly at most of the order $(few\,hundred\,eV)^2/(1 \,TeV)
\approx 10^{-8}\,eV$. As a result, the bulk of the neutrino mass, in
this scenario, is {\em Dirac}. Furthermore, if the unconventional fermions
are not too heavy, say lighter than $500\, GeV$, nor too light,
i.e. heavier than the top quark, the neutrino mass scales vary between 
$1\,eV$ and approximately $0.07\,eV$. From Table I, one can tentatively
conclude that if the unconventional fermions are heavy, i.e.
with masses ranging from $300\,GeV$ to $500\,GeV$, the neutrino mass
scales will range between $0.2\,eV$ and $0.7\,eV$. This would imply
that one might have a situation in which neutrinos are nearly
degenerate in order to satisfy the oscillation data. If, on the
other hand, the unconventional fermions were to be lighter, i.e.
with masses ranging from $181\,GeV$ to $400\,GeV$, one could
have a scenario in which the neutrino mass matrix is hierarchical.

\item Table II:

This is the extreme case of democratic mass matrices for the normal
fermions.

If the masses of the unconventional fermions were to lie between $181\,
GeV$ and $400\,GeV$, the neutrino mass scales would be of the order
of a few $eV$s or more. In light of cosmological constraints as
well as of oscillation data, this particular case might even be ruled
out. It is amusing to note that this scenario of extreme democratic
mass matrices for the normal fermions does not work but itself,
regardless of the neutrino sector, because it cannot reproduce the
correct mass spectrum and the CKM matrix.

\item Intermediate cases:

In between the above two bounds, there are models, e.g.
(\cite{HS,HSS}), which deviate somewhat from a pure
democratic mass matrix model but which can fit
fairly well the mass spectrum as well as the CKM matrix.
In this model, the mass scales are are roughly half the value
of the largest mass eigenvalues. In rescaling Table I by
a factor of 1/2, one notices that, in order for the lightest
unconventional fermion to be heavier than the top quark, the
smallest neutrino mass scale is around $0.2-0.4\,eV$. This implies
that neutrinos are nearly degenerate.

\ei

Although a more extensive investigation of the above questions
for various scenarios of mass matrices is warranted-a subject
of a next paper- a preliminary conclusion can be drawn from
the above results. From the consideration of the lower bound
on the lightest unconventional fermion, it appears that neutrinos
in our scenario are more likely to be near-degenerate with mass
lying around a few tenths of an $eV$. This would imply that
mixing angles as deduced from various oscillation data mainly
come from the charged lepton sector.

\section{Summary}

We have presented, in this paper, a model of quark-lepton
mass unification which ``marries'' two TeV-scale scenarios:
Early Unification of Quarks and Leptons \cite{HBB,BH2,BH3} and
Large Extra Dimensions \cite{AHS,ref6,H2} (as applied to
neutrino masses), has been presented. Explicitely, the
early unification model $G_{PUT} =
SU(4)_{\rm PS}\otimes SU(2)_L \otimes SU(2)_R \otimes SU(2)_H$
is embedded in 4+1 dimensions with the extra spatial dimension, $y$,
being compactified on an orbifold $S_{1}/Z_{2}$. Chiral zero modes
are localized along the extra spatial dimension by kinks that
come from two background scalar fields, one of which transforms 
non-trivially under $G_{PUT}$. Additional scalars are used to
break $G_{PUT}$ down to the SM. The model contains the following
features.

\bi

\item The breaking
of $SU(4)_{PS}$ splits the positions, along $y$, of wave functions
of the zero modes of ``quarks'' and ``leptons''.

\item The breaking
of $SU(2)_R$ gives rise to two vastly diferent profiles for
the wave functions of the ``right-handed''
zero modes.

\item Since a $SU(2)_H$ doublet groups together
a conventional quark (or lepton) with an unconventional one,
the breaking of $SU(2)_H$ splits the locations, along $y$,
of the wave functions of the conventional fermions relative
to those of the unconventional ones.

\item The breaking of $SU(2)_L \otimes U(1)_Y$ provides a mass 
scale for all the fermions.

\item The size of the effective Yukawa couplings, which depends
on the overlap between right- and left-handed wave functions, is
characterized by the separation along the extra dimension
$y$ between these two wave functions,
$\Delta y$. In this model, we were able to show that there is 
a relationship between the quark separation, 
$\Delta y^{(q)}$ and the lepton separation $\Delta y^{(l)}$, 
and also that of the unconventional fermions. This is due to
the early unification scenario discussed in \cite{BH2,BH3}
and again in this paper. It translates into relationships
between mass scales that appeared in fermion mass matrices and
are valid in the TeV range. This is what is referred to
as {\em early quark-lepton mass unification} in our scenario. A summary
of its ramification is listed below.

\item A feature of the
$SU(4)_{\rm PS}\otimes SU(2)_L \otimes SU(2)_R \otimes SU(2)_H$ model
is the existence of ``quarks'' and ``leptons'' with unconventional
electric charges. As described in the model, these unconventional
fermions acquire masses from the same sources as conventional
fermions. Because of the existence of relationships between
mass scales of different sectors, a consequence of early quark-lepton
unification, we have found a strong correlation between the
Dirac masses of the neutrinos and those of the unconventional
fermions: The neutrino Dirac masses increase the heavier the
unconventional fermions become.
For example, by requiring that the masses of these 
unconventional fermions lie between the top mass and $1\,TeV$
(the lower bound is experimentally obvious while the
upper bound refers more to the wish of not having a strong
Yukawa coupling regime), we
found that the mass scales of the neutrino sector range from
approximately $0.07\,eV$ to roughly $80\,eV$, as can be seen
in Tables I and II. The Dirac masses of the neutrinos are
{\em naturally small} in this scenario. Any Majorana
contribution to the total mass through the see-saw mechanism
would have to be {\em negligible}.

From the above arguments, our scenario accomodates
{\em naturally light Dirac neutrinos} without having to use
the see-saw mechanism. Since there is a correlation between
neutrino masses and those of the unconventionally charged
fermions in the 
$SU(4)_{\rm PS}\otimes SU(2)_L \otimes SU(2)_R \otimes SU(2)_H$ model,
a light neutrino of mass less than $1\,eV$ (as a concordance of data
seems to ``indicate'' \cite{neutrino}) will imply that the 
unconventionally charged
fermions are not too heavy (see Tables I and II) and there
might be a chance to observe them, if they exist, at future colliders
\cite{FHS}.

\ei

\begin{acknowledgments}
I would like to thank Lia Pancheri and Gino Isidori for the warm
hospitality in the Theory Group at Frascati.
This work is supported in parts by the US Department
of Energy under grant No. DE-A505-89ER40518. 
\end{acknowledgments}


\begin{thebibliography}{99}
\bibitem{BEGN}
A.~J.~Buras, J.~R.~Ellis, M.~K.~Gaillard and D.~V.~Nanopoulos,
Nucl.\ Phys.\ B {\bf 135}, 66 (1978).
\bibitem{GUT}
H.~Georgi and S.~L.~Glashow,
Phys.\ Rev.\ Lett.\  {\bf 32}, 438 (1974);

H.~Georgi, H.~R.~Quinn and S.~Weinberg,
Phys.\ Rev.\ Lett.\  {\bf 33}, 451 (1974).
\bibitem{HBB}
P. Q. Hung, A. J. Buras and, J. D. Bjorken, Phys. Rev. D {\bf 25},
805 (1982).
\bibitem{BH2}
A.~J.~Buras and P.~Q.~Hung,
Phys.\ Rev.\ D {\bf 68}, 035015 (2003).
\bibitem{BH3}
A.~J.~Buras, P.~Q.~Hung, Ngoc-Khanh ~Tran, A. ~Poschenrieder, E. ~Wyszomirski,
Nucl.\ Phys.\ B {\bf 699}, 253 (2004).
\bibitem{H1}
P.~Q.~Hung, Proceedings of
Coral Gables Conference on Launching of Belle Epoque in High-Energy 
Physics and Cosmology (CG 2003), Ft. Lauderdale, Florida, Dec 17-21 (2003).
\bibitem{AHS}
N.~Arkani-Hamed, S.~Dimopoulos and G.~R.~Dvali,
Phys.\ Lett.\ B {\bf 429}, 263 (1998)
[arXiv:hep-ph/9803315];
I.~Antoniadis, N.~Arkani-Hamed, S.~Dimopoulos and G.~R.~Dvali,
Phys.\ Lett.\ B {\bf 436}, 257 (1998)
[arXiv:hep-ph/9804398]
\bibitem{ref6}
Nina Arkami-Hamed, S. Dimopoulos, G. R. Dvali,
and John March-Russell, Phys.\ Rev.\ D {\bf 65}, 024032 (2002);
K. Dienes, Emilian Dudas, and Tony Ghergetta, 
Nucl.\ Phys. {\bf B557}, 25 (1999);
J.~M.~Frere, G.~Moreau, E.~Nezri,
Phys.\ Rev.\ D {\bf 69}, 033003 (2003).
\bibitem{H2}
P.~Q.~Hung,
Phys.\ Rev.\ D {\bf 67}, 095011 (2003).
\bibitem{grossman}
See e.g. Y.~Grossman, R.~Harnik, G.~Perez, M.~D.~Schwartz and Z.~Surujon,
arXiv:hep-ph/0407260, and references therein.
\bibitem{HS}
P.~Q.~Hung and M.~Seco,
Nucl.\ Phys.\ B {\bf 653}, 123 (2003).
\bibitem{HSS}
P.~Q.~Hung, M.~Seco and A.~Soddu,
Nucl.\ Phys.\ B {\bf 692}, 83 (2004).
\bibitem{PS}
J.~C.~Pati and A.~Salam,
Phys.\ Rev.\ D {\bf 10}, 275 (1974).
\bibitem{seesaw}
M. Gell-Mann, P. Ramond, and R. Slansky in Supergravity,
edited by D. Freedman et al. (1979); T. Yanagida, KEK lectures (unpublished)
(1979); R. N. Mohapatra and G. Senjanov\'{i}c, Phys. Rev. Lett. {\bf 44}, 912
(1980).
\bibitem{smirnov}
A.~Y.~Smirnov,
Talk given at SEESAW25: International Conference on the Seesaw Mechanism and the Neutrino Mass, Paris, France, 10-11 Jun 2004, hep-ph/0411194.
\bibitem{CHP}
Z.~Chako, L.~J.~Hall and M.~Perelstein,
JHEP\ {\bf 0301}, 001 (2003).
\bibitem{democratic}
See e.g.
P.~Kaus and S.~Meshkov, Phys.\ Rev.\ D {\bf 42}, 1863 (1990), and
references therein.
\bibitem{hierarchical}
There is a long history on this subject. Below is an incomplete
set of references.
H.~Fritzsch,
Phys.\ Lett.\ B {\bf 73}, 317 (1978);
P.~Ramond, R.~G.~Roberts and G.~G.~Ross,
Nucl.\ Phys.\ B {\bf 406}, 19 (1993)
[arXiv:hep-ph/9303320];
D.~s.~Du and Z.~z.~Xing,
Phys.\ Rev.\ D {\bf 48}, 2349 (1993);
L.~J.~Hall and A.~Rasin,
Phys.\ Lett.\ B {\bf 315}, 164 (1993)
[arXiv:hep-ph/9303303];
H.~Fritzsch and D.~Holtmannspotter,
Phys.\ Lett.\ B {\bf 338}, 290 (1994)
[arXiv:hep-ph/9406241];
P.~S.~Gill and M.~Gupta,
J.\ Phys.\ G {\bf 21}, 1 (1995).
P.~S.~Gill and M.~Gupta,
Phys.\ Rev.\ D {\bf 56}, 3143 (1997)
[arXiv:hep-ph/9707445];
H.~Lehmann, C.~Newton and T.~T.~Wu,
Phys.\ Lett.\ B {\bf 384}, 249 (1996);
Z.~z.~Xing,
J.\ Phys.\ G {\bf 23}, 1563 (1997)
[arXiv:hep-ph/9609204];
K.~Kang and S.~K.~Kang,
Phys.\ Rev.\ D {\bf 56}, 1511 (1997)
[arXiv:hep-ph/9704253];
T.~Kobayashi and Z.~z.~Xing,
Mod.\ Phys.\ Lett.\ A {\bf 12}, 561 (1997)
[arXiv:hep-ph/9609486];
T.~Kobayashi and Z.~z.~Xing,
Int.\ J.\ Mod.\ Phys.\ A {\bf 13}, 2201 (1998)
[arXiv:hep-ph/9712432];
J.~L.~Chkareuli and C.~D.~Froggatt,
Phys.\ Lett.\ B {\bf 450}, 158 (1999)
[arXiv:hep-ph/9812499];
J.~L.~Chkareuli, C.~D.~Froggatt and H.~B.~Nielsen,
Nucl.\ Phys.\ B {\bf 626}, 307 (2002)
[arXiv:hep-ph/0109156];
A.~Mondragon and E.~Rodriguez-Jauregui,
 ``The breaking of the flavor permutational symmetry: Mass textures and  the
%
Phys.\ Rev.\ D {\bf 59}, 093009 (1999)
[arXiv:hep-ph/9902240];
H.~Nishiura, K.~Matsuda and T.~Fukuyama,
Phys.\ Rev.\ D {\bf 60}, 013006 (1999)
[arXiv:hep-ph/9902385];
G.~C.~Branco, D.~Emmanuel-Costa and R.~Gonzalez Felipe,
Phys.\ Lett.\ B {\bf 477}, 147 (2000)
[arXiv:hep-ph/9911418];
R.~Rosenfeld and J.~L.~Rosner,
 ``Hierarchy and anarchy in quark mass matrices, or can hierarchy tolerate
%
Phys.\ Lett.\ B {\bf 516}, 408 (2001)
[arXiv:hep-ph/0106335];
H.~Fritzsch and Z.~z.~Xing,
Phys.\ Lett.\ B {\bf 555}, 63 (2003)
[arXiv:hep-ph/0212195].
\bibitem{PPM}
G.~C.~Branco, J.~I.~Silva-Marcos and M.~N.~Rebelo,
Phys.\ Lett.\ B {\bf 237} 446 (1990); G.~C.~Branco and J.~I.~Silva-Marcos,
Phys.\ Lett.\ B {\bf 359}, 166 (1995); G.~C.~Branco, D.~Emmanuel-Costa and 
J.~I.~Silva-Marcos, Phys.\ Rev.\ D {\bf 56}, 107 (1997).
\bibitem{neutrino}
For the latest, see e.g. S.~Hannestad,
hep-ph/0412181;
J.~A.~Aguilar-Saavedra, G.~C.~Branco and F.~R.~Joaquim,
Phys.\ Rev.\ D {\bf 69}, 073004 (2004)
[arXiv:hep-ph/0310305];
V.~Barger, D.~Marfatia and K.~Whisnant,
Int.\ J.\ Mod.\ Phys.\ E {\bf 12}, 569 (2003)
[arXiv:hep-ph/0308123].
\bibitem{FHS}
See e.g. P.~Frampton, P.~Q.~Hung, M.~Sher,
Phys. Rept. {\bf 330}, 263 (2000)
[arXiv:hep-ph/9903387]
\end{thebibliography}
\end{document}